\documentclass[twocolumn,aps,floatfix,superscriptaddress,longbibliography]{revtex4-1}

\pdfoutput=1 %This is for the ArXiv submission only
\usepackage[colorlinks=true, urlcolor=blue, anchorcolor=blue, citecolor=blue,filecolor=blue,linkcolor=blue,menucolor=blue]{hyperref}

\usepackage{amsmath,amssymb,eucal,graphicx,float,epstopdf}
\usepackage{subfigure}

\def\hatmz{\mu}
\def\hatmc{\mu}

\begin{document}

\title{Dynamics of feedback Ising model}

\author{Yi-Ping Ma}
\email{yiping.ma@northumbria.ac.uk}
\affiliation{Department of Mathematics, Physics and Electrical Engineering, Northumbria University, Newcastle upon Tyne, NE1 8ST, UK}

\author{Ivan Sudakow}
\email{ivan.sudakow@open.ac.uk}
\affiliation{School of Mathematics and Statistics, The Open University, Milton Keynes, MK7 6AA, UK}

\author{P. L. Krapivsky}
\email{pkrapivsky@gmail.com}
\affiliation{Department of Physics, Boston University, Boston, Massachusetts 02215, USA}
\affiliation{Santa Fe Institute, Santa Fe, New Mexico 87501, USA}

\author{Sergey A. Vakulenko}
\affiliation{Institute of Problems in Mechanical Engineering, Russian Academy of Sciences, Bolshoj pr., 61, St.\,Petersburg, 199178, Russia}

\begin{abstract}
We study the dynamics of a mean-field Ising model whose coupling depends on the magnetization via a linear feedback function. A key feature of this linear feedback Ising model (FIM) is the possibility of temperature-induced bistability, where a temperature increase can favor bistability between two phases. We show that the linear FIM provides a minimal model for a transcritical bifurcation as the temperature varies. Moreover, there can be two or three critical temperatures when the external magnetic field is non-negative. In the bistable region, we identify a Maxwell temperature where the two phases are equally probable, and we find that increasing the temperature favors the lower phase. We show that the probability distribution becomes non-Gaussian on certain time intervals when the magnetization converges algebraically at either zero temperature or critical temperatures. Near critical points in the parameter space, we derive a Fokker-Planck equation, construct the families of equilibrium distributions, and formulate scaling laws for transition rates between two stable equilibria. The linear FIM offers considerable flexibility in controlling steady-state bifurcations and their associated equilibrium distributions, which can be desirable for modeling feedback systems across various disciplines.
\end{abstract}

\maketitle

\section{Introduction}

The Ising model (IM), proposed as a toy model of ferromagnetism \cite{Lenz20, Ising25}, acquired a paradigmatic status \cite{Baxter82, Itzykson} and remains the go-to framework for explaining how simple, local interactions give rise to collective order and phase transitions in complex systems. In the classical version of the IM, the interaction strengths are fixed; the magnetic field is also imposed externally, albeit it can vary with time. Therefore, the spins do not influence the parameters that govern their behavior. Real-world complex systems are rarely so one-way. Ecological populations can modify their habitats. For instance, forests influence soil composition and local climate. Regional geography can also affect global climate, e.g., sea ice cover alters the reflectivity of the planet. In social systems, agents continually reshape the information environment they inhabit. These kinds of {\em feedback loops}, in which a macroscopic observable feeds back into the microscopic rules, can dramatically change the stability of states and transitions between states.

Far from being limited to magnetic spins, the Ising paradigm has been extended to model diverse phenomena across disciplines, wherever entities have binary-like states and local interactions. Many complex systems have been mapped onto or understood via Ising-like models. In each case, spin-up/spin-down corresponds to two alternative states, and collective phases represent coordinated behavior or consensus emerging from pairwise interactions. 

In neuroscience, IMs have been used to describe neural populations by treating each neuron as a spin that is either firing or silent. For instance, Schneidman et al.~\cite{Schneidman2006Weak} showed that even weak pairwise correlations between neurons are enough to explain strongly collective firing patterns using an Ising-like model. The success of this binary-state abstraction indicates that neural circuits hover near criticality and that low-order couplings can reconstruct large-scale information processing.

IMs have gained popularity in the modeling of ecology and climate. On planar lattices, predator-prey Lotka-Volterra models exhibit transitions from coexistence to extinction resembling the behavior of an Ising ferromagnet crossing the critical point \cite{Matsuda1992_LatticeLV}. At cryospheric scales, a two-dimensional ice-water spin model reproduces the size distribution and fractal geometry of Arctic melt ponds without invoking large-scale climate tipping \cite{Ma2019}. A lattice model of tropical convection  \cite{Majda2012} maps active/inactive sites onto spins whose flip rates depend on a slow moisture field, thereby yielding a link between microphysics and planetary variability.

In socio-economic systems, binary spins imitate agents on social networks and interactions between neighbors via voter model or Glauber rules may generate consensus, polarization, and critical slowing, thereby mimicking social agreement, stalemate, and tipping \cite{Castellano2009_RevModPhys}. In finance, coupling traders' buy/sell spins through a market-wide magnetic field proportional to excess demand captures bubbles, crashes, and heavy-tailed returns within the same Ising vocabulary \cite{Kaizoji2000_PhysicaA}.

The above examples show that allowing large-scale system properties to influence local interactions, like in a {\it feedback Ising model} (FIM) analyzed in this paper, offers a simple way to explain complex behaviors such as hysteresis, multi-step transitions, and stable intermediate states in many fields. Most extensions of the IM treat the macroscopic variables, e.g., external magnetic field, temperature, coupling strength, network topology, etc., as constant, externally controlled parameters. Real systems, however, often feature endogenous feedback: the microscopic configuration modifies macroscopic variable(s) that feed back onto the microscopic dynamics. Capturing such co-evolution requires letting the Hamiltonian depend on collective observables.

Prior attempts to construct a feedback loop in the IM typically incorporate feedback into the external field or the temperature. Bornholdt's market model reproduces market observations by letting the local external field depend on both the local spin and the global magnetization \cite{Bornholdt2001}, while a later variant lets the ``social temperature'' depend on the global magnetization \cite{krause2012opinion}. In a different context, linear feedback endows mean-field Landau theory with an Andronov-Hopf bifurcation, which implies the emergence of limit cycles in a fully connected Ising system \cite{DeMartino2019}. In Ref.~\cite{ma2025mixed}, we have introduced the general FIM whose coupling is an arbitrary function of the magnetization. At zero temperature, the FIM can describe binary phase transformations with an arbitrary bifurcation diagram, thereby establishing the relevance of feedback coupling to data-driven modeling. The simplest FIM with linear feedback coupling already exhibits many salient features of the general FIM and  should find most applications.

Here, we analyze the dynamics of the simplest FIM whose coupling grows linearly with the magnetization. The general FIM keeps the microscopic degrees of freedom of the standard mean-field Ising set-up, viz., spins $s_i=\pm1$, on a complete graph of  $N$ sites, but additionally lets the two-spin coupling depend on the instantaneous magnetization
\begin{equation}
\label{m:def}
m = \frac{1}{N}\sum_{k=1}^{N} s_k\,.
\end{equation}
We limit ourselves to linear dependence of the coupling on $m$, yielding the linear FIM with the Hamiltonian
\begin{equation}
\label{ham-F} 
\mathcal H_{\mathrm{FIM}}
   \;=\;
   -h\sum_{i=1}^{N} s_i
   \;-\;
   \frac{1}{N}\bigl[1+\gamma m\bigr]
   \sum_{i<j} s_i s_j.
\end{equation}
Here $h$ is an external field. The feedback parameter $\gamma$ measures the strength of the feedback. Although positive $\gamma$ reinforces the majority orientation, while negative $\gamma$ weakens it, these two scenarios are equivalent by flipping the signs of $s_i$, $h$, and $\gamma$ in the Hamiltonian. This observation allows us to pose $\gamma>0$.

Our goal is to reveal the non-equilibrium properties of the FIM via {\it Glauber dynamics}, although we note that the equilibrium properties of related systems have been studied in various settings. The feedback coupling in the general FIM, e.g., $[1+\gamma m]$ in the linear FIM, can be written as a Taylor series in $m$. Thus, the Hamiltonian is a weighted sum of $p$-spin interactions, $p\in\mathbb{N}$, which are called equivalent-neighbor interactions or, more generally, separable interactions in early literature \cite{Capel1976,Ouden1976}. For ferromagnetic quadratic interactions, Bogolyubov Jr.~\cite{Bogolyubov1972} used a minimax principle to derive the free energy. Den Ouden, Capel, Perk, et al.~then extended this method to higher-degree separable interactions \cite{Capel1976} and more generally separable functions of short-range interactions \cite{Ouden1976}. This exact result can be simplified using a convex-envelope construction, i.e., a Legendre transform \cite{Capel1977}. In this framework, the free energy is first recast from the fixed-field (canonical) ensemble, e.g., fixed-$h$, to the fixed-density ensemble, e.g., fixed-$m$, and then subject to a generalized Maxwell construction. This framework can reveal the critical behavior of systems with short-range interactions under small perturbations, including critical-exponent renormalization and first-order transitions \cite{Capel1979}.

Beyond this framework, $p$-spin models have attracted much interest in both theories and applications. Consider first 1-spin interactions, which describe the response of the magnetization $m$ to the external field $h$. This is typically captured by the term $-hm$ in the Hamiltonian, e.g., Eq.~(\ref{ham-F}), to guarantee the fundamental property of magnets that $m\to\pm1$ as $h\to\pm\infty$. This term also guarantees the applicability of the Legendre transform with the conjugate variables $h$ and $m$. However, we note that 1-spin interactions are sometimes omitted to focus on the intrinsic properties of the system.

Beyond 1-spin interactions, $p$-spin models often replace 2-spin interactions by $p$-spin interactions with $p\geq3$, e.g., in spin glasses \cite{mezard2009information} and in the Sachdev-Ye-Kitaev model for interacting Majorana fermions \cite{sachdev1993gapless,kitaev2018soft}. The fully connected $p$-spin model inspired several algorithms including reverse annealing for combinatorial optimization \cite{ohkuwa2018reverse,yamashiro2019dynamics} and dense associative memory for machine learning \cite{krotov2016dense}. In the $p$-spin Curie-Weiss model, the limiting distribution of the magnetization can be a discrete mixture with either two or three atoms \cite{mukherjee2021fluctuations}, which then implies phase transitions of the maximum likelihood estimators \cite{MukherjeeSonBhattacharya2025}.

There are several studies on $p$-spin models with multiple $p$'s. These so-called mixed $p$-spin models often keep 2-spin interactions of the classical IM but add either 3-spin or 4-spin interactions. In early studies, 4-spin interactions arise from spin-phonon interactions in the compressible IM \cite{BSLee_1971}, which motivates an exactly solved model with a nearest-neighbor pair (2-spin) interaction and an infinite-range pair-pair (4-spin) interaction \cite{oitmaa1975critical}. The equilibrium properties of the mean-field IM with 2-spin and 4-spin interactions \cite{thompson1974model,Capel1976}, possibly with $2n$-spin interactions for $n\geq3$ \cite{mckerrell1972critical,bowers1972first}, are also analyzed.

In contrast, early interest in 3-spin interactions largely relates to the well-known exact solution of the Baxter-Wu model \cite{Baxter_Wu}, i.e., the IM with 3-spin interactions on the triangular lattice. This model with additional 2-spin interactions cannot be solved exactly; its mean-field approximation has been formulated in Ref.~\cite{Hemmer}. Recently, the mean-field IM with 2-spin and 3-spin interactions has been applied to human-AI ecosystems \cite{AI-Human}, where the 3-spin interactions describe two humans interacting with an AI agent or two AI agents interacting with a human. The equilibrium properties of this model are further analyzed \cite{CW-23,CW-3}. Although some of our results are known from these studies, our FIM unifies mixed $p$-spin models via feedback couplings and thus provides a general toolbox to model feedback systems. Moreover, our focus on non-equilibrium properties via Glauber dynamics aligns with popular modeling techniques from nonlinear dynamics.

This paper is organized as follows. In Sec.~\ref{sec:CW}, we revisit the Curie-Weiss model with Glauber single spin-flip dynamics. We examine both small fluctuations around an equilibrium and thermally activated switching between two equilibria. Section \ref{FIM1} develops the FIM in full, revealing its unique features, including temperature-induced bistability on the phase diagram and the dynamics of probability distributions at zero temperature. In Sec.~\ref{LTR}, the master equation for Glauber dynamics is reduced to a Fokker-Planck (FP) equation near critical points in the parameter space, yielding asymptotic formulas for equilibrium distributions and scaling laws for transition rates. In Sec.~\ref{sec:disc}, we briefly discuss how such feedback-coupled interactions illuminate phenomena across magnetic, ecological, climatic, and socio-economic systems.

\section{Curie-Weiss Model and Glauber dynamics} 
\label{sec:CW}

A classical system mimicking a phase transition goes back to Curie \cite{Curie} and Weiss \cite{Weiss}, who tried to understand the phenomenon of spontaneous magnetization and introduced what is now known as the mean-field approximation. The precise setting where such approximations are exact is the IM on a complete graph, where every spin interacts with every other spin. This model is now called the Curie-Weiss model \cite{Kac}. This simplest exactly solvable model occupies a key place in science \cite{thompson2015mathematical,friedli2017statistical}, and even its equilibrium characteristics continue to attract attention \cite{CW-rev, CW-25}. Endowing the Curie-Weiss model with Glauber spin-flip dynamics gives the simplest exactly solvable {\em dynamical} model, shedding light on the dynamics of the IM in finite dimensions. 

Here we consider the dynamical Curie-Weiss model, with or without an external magnetic field. Moreover, we derive the equilibrium distribution for general mean-field IMs and use this distribution to calculate transition times between two stable equilibria. A few of our results could be new, or at least hard to find in the literature, but our chief goal is to present a framework that we apply to the FIM in Sec.~\ref{FIM1}. 

\subsection{Zero external magnetic field}
\label{sec:im-zero}

For the one-dimensional IM, Glauber proposed \cite{Glauber63} a simple spin-flip rate $w_i(s_i\to -s_i)$ that agrees with the detailed balance. The Glauber rate admits a straightforward generalization to an IM on an arbitrary simple graph \cite{Diestel}, i.e., an undirected graph without loops and multiple edges. The Hamiltonian of such an IM reads 
\begin{equation}
\label{ham-gen}
\mathcal{H}_\text{IM}=-\sum_{\langle i j\rangle} J_{ij} s_i s_j
\end{equation}
where the sum is over all edges $\langle i j\rangle$ and couplings $J_{ij}$ are arbitrary. The Glauber spin-flip rate has the form 
\begin{equation}
\label{glauber-rate}
w_i=\frac{1}{2}\left(1-\tanh\left[\beta s_i\sum_{j\in [i]} J_{ij} s_j\right]\right)
\end{equation}
where the sum runs over all neighbors of vertex $i$. The spin-flip rate \eqref{glauber-rate} is the simplest one that agrees with the detailed balance; other rates with such a property exist but are analytically intractable already in one dimension \cite{KRB}. On the hypercubic lattices, the ferromagnetic IM endowed with Glauber dynamics has been solved only in one dimension \cite{Glauber63,Oppenheim,Felderhof}, although a lot is also known about qualitative behaviors of higher-dimensional Ising-Glauber models  \cite{KRB,Bray94}.

In the ferromagnetic IM on the complete graph, the Curie-Weiss model, the couplings are equal, positive, and should scale inversely proportional to the number of vertices to ensure the proper thermodynamic behavior, e.g., the extensivity of energy, in the $N\to\infty$ limit. We thus set $J_{ij}=N^{-1}$, and the Hamiltonian \eqref{ham-gen} becomes
\begin{equation}
\label{hamiltonian}
\mathcal{H}_\text{CW}=-\frac{1}{N}\sum_{i<j} s_i s_j
\end{equation}
while the spin-flip rate \eqref{glauber-rate} simplifies to 
\begin{equation}
\label{glauber-mf}
w_i=\frac{1}{2}\,(1-s_i\tanh \beta m)
\end{equation}
where $\beta=1/T$ is the inverse temperature. Using the general relation 
\begin{equation}
\label{si}
\frac{d \langle s_i\rangle}{dt}= -2 \langle s_i w_i\rangle
\end{equation}
and summing over $i$ one derives an ordinary differential equation (ODE) for the magnetization 
\begin{equation}
\label{m}
\frac{dm}{dt}= -m + \tanh \beta m.
\end{equation}

More precisely, one obtains 
\begin{equation}
\label{m-av}
\frac{d \langle m\rangle}{dt} = -  \langle m\rangle +  \langle\tanh(\beta m)\rangle
\end{equation}
which is not even an equation for the average magnetization since $\langle\tanh(\beta m)\rangle \ne \tanh(\beta \langle m\rangle)$ for random magnetization. Only the infinite system limit, the magnetization becomes deterministic and we can use Eq.~\eqref{m}. To avoid cluttering formulas, we write $m$ instead of $\langle m\rangle$ even for finite systems. 

The phase transition occurs at $\beta_c=1$. For $\beta\leq \beta_c=1$, the equilibrium solution is $m=0$. For $\beta>\beta_c$, in addition to the zero-magnetization solution, there are non-trivial solutions $m=\pm m_\infty$ where $m_\infty$ is the positive solution of $m_\infty=\tanh \beta m_\infty$. In the proximity of the critical temperature $m_\infty\to\sqrt{3(\beta-\beta_c)}$.  The zero-magnetization solution is unstable, while other equilibria are stable. The existence of the two stable solutions is known as spontaneous symmetry breaking.

When $\beta\leq \beta_c=1$, the magnetization decays to zero. To study the asymptotic behavior near $m=0$, one expands $\tanh \beta m$ and gets
\begin{equation}
\label{m-exp}
\frac{dm}{dt}= -(1-\beta)m - \frac{\beta^3 m^3}{3}
\end{equation}
after dropping higher order terms. When $\beta < \beta_c=1$, one neglects the cubic term and obtains $m\sim e^{-(1-\beta)t}$ showing that the magnetization decreases exponentially.

Near criticality ($\beta\approx\beta_c$), Eq.~(\ref{m-exp}) agrees with the normal form of pitchfork bifurcation \cite{strogatz2024nonlinear}. At criticality,  $\dot m=-m^3/3$ leading to an algebraic decay
\begin{equation}
\label{m-at}
m=\text{sgn}(m_0) \sqrt{\frac{3}{2t}} 
\end{equation}
Therefore the critical dynamic exponent is $z_c=2$. 

Below the critical temperature ($\beta>\beta_c$), the magnetization approaches $\pm m_\infty$. From Eq.~(\ref{m}) we find that the
approach to final magnetization is exponential: 
\begin{equation}
\label{m-below}
1-\frac{m}{m_\infty}\sim e^{-\Lambda t}\,, 
\quad \Lambda=1-\frac{\beta}{\cosh^2(\beta m_\infty)}
\end{equation}
In the proximity of the critical temperature, the inverse relaxation time is $\Lambda\simeq 2(\beta-\beta_c)$.

The above analysis applies to infinite systems. Let us now turn to systems with large but finite $N$. The zero-temperature dynamics drive the system to a ground state with fully aligned spins. If the initial magnetization $m_0$ is positive, it is driven to the up phase, i.e., the number $M$ of up spins increases. Let  $P_M(t)$ be the probability of having $M$ up spins and $N-M$ down spins. At zero temperature, the probabilities $P_M(t)$ evolve according to
\begin{equation}
\label{PM} 
\frac{d P_M}{dt}=(N-M+1)P_{M-1}-(N-M)P_M.
\end{equation}
We have $P_M\to \delta_{M,N}$ as $t\to\infty$. For finite time, the probability $P_M(t)$ approaches a Gaussian form
\begin{equation}\label{eq:PM-gauss}
P_M(t)=\frac{1}{\sqrt{2\pi\Delta^2}}\,
\exp\left[-\frac{(M-\langle M\rangle)^2}{2\Delta^2}\right]
\end{equation}
with
\begin{subequations}
\begin{align}
& \langle M\rangle=N\left[1-\frac{1-m_0}{2}\,e^{-t}\right],
\label{Mav} \\
& \Delta^2=N\,\frac{1-m_0}{2}\left(e^{-t}-e^{-2t}\right).
\label{Dav} 
\end{align}
\end{subequations}
The Gaussian shape can be justified by utilizing the $1/N$ expansion. Furthermore, Eqs.~(\ref{Mav}--\ref{Dav}) can be derived from the master equation (\ref{PM}). The time evolution of the mean $\langle M\rangle$ in Eq.~(\ref{Mav}) agrees with the solution of Eq.~(\ref{m}) with initial condition $m(t=0)=m_0$ at $\beta=\infty$. The variance $\Delta^2$ in Eq.~(\ref{Dav}) increases from 0 at $t=0$, peaks at $t=\log2$, and decreases to 0 as $t\to\infty$.

Let us now look at the critical dynamics ($\beta=\beta_c=1$). Here, we write the master equation for the probabilities $P_M(t)$ applicable to general mean-field IMs:
\begin{align} 
\frac{d P_M}{dt}&=W_-(M+1)\,P_{M+1}+W_+(M-1)\,P_{M-1}\nonumber\\
&-[W_-(M)+W_+(M)]\,P_M,
\label{MasterEq}
\end{align}
where $W_\pm(M)$ denote the flipping rates for $M\to M\pm 1$. At criticality, these rates are obtained from Eq.~\eqref{glauber-mf} and the relation $m=\frac{2M}{N}-1$:
\begin{eqnarray*}
W_-(M)&=&\frac{1}{2}\,M\left[1-\tanh\left(\frac{2M}{N}-1\right)\right],\\
W_+(M)&=&\frac{1}{2}\,(N-M)\left[1+\tanh\left(\frac{2M}{N}-1\right)\right].
\label{transM}
\end{eqnarray*}
Using either the master equation (\ref{MasterEq}) or the Boltzmann distribution $P_M(\infty)\propto 2^{-N}{N\choose M} e^{-{\cal H}}$, one can derive the equilibrium ($t=\infty$) distribution $P_{\rm eq}(m)\equiv P_M(\infty)$:
\begin{subequations}
\label{crit-N-CW}
\begin{align}
\label{crit-inf}
&P_{\rm eq}(m)\to \frac{N^{1/4}}{C}\,\exp\!\left[-\frac{Nm^4}{12}\right] \\
\label{crit-C}
&C=(3/4)^{1/4}\Gamma(1/4)= 3.374\,010\,197\,800 \ldots 
\end{align}
\end{subequations}
where the amplitude $C$ is fixed by the normalization requirement; see also Refs.~\cite{ellis1978limit,papangelou1989gaussian} for rigorous derivations.

The phase transition point in the infinite system is replaced by a finite region when the system size $N$ is finite \cite{Stanley}. The width of this region, known as a `scaling window', vanishes as $N\to\infty$. For the Curie-Weiss model, the width of the scaling window scales as $N^{-1/2}$. To appreciate the behavior in the scaling window, one sets $\beta=1-b/\sqrt{N}$ with $b=O(1)$. The equilibrium distribution \eqref{crit-inf} corresponds to $b=0$. One can generalize it \cite{CW-25} to arbitrary finite $b$ and obtain the distribution
\begin{equation}
\label{crit-window}
P_{\rm eq}(m)\to \frac{N^{1/4}}{C(b)}\,\exp\!\left[-b\,\frac{\sqrt{N}\,m^2}{2}-\frac{Nm^4}{12}\right]
\end{equation}
valid across the scaling window. The methods that we used in deriving \eqref{crit-inf} can be also adjusted for the derivation of \eqref{crit-window}.

The amplitude $C(b)$ in Eq.~\eqref{crit-window} is fixed by normalization. The results differ depending on whether $b$ is positive or negative. In the former case, the amplitude can be expressed via the modified Bessel function of the second kind; for $b<0$, the amplitude admits an expression via the modified Bessel functions of the first kind:
\begin{equation}
C(b) = \frac{\pi \sqrt{3|b|}}{2}\, e^B\times  
\begin{cases}
\frac{\sqrt{2}}{\pi} \,K_{\frac{1}{4}}(B) & b>0\\
 I_{-\frac{1}{4}}(B)+I_{\frac{1}{4}}(B)  & b<0
 \end{cases}
\end{equation}
where $B=3b^2/8$.

The equilibrium distribution \eqref{crit-inf} at the critical point is non-Gaussian, while the distribution at zero temperature, Eq.~(\ref{eq:PM-gauss}), is Gaussian throughout the evolution. We explore the evolution at the critical point using the FP equation for $P_M(t)$ derived in Appendix \ref{sec:conv-eq-zero}. In the $t\to\infty$ limit, the solution of the FP equation is \eqref{crit-inf}; this provides an alternative derivation of the equilibrium distribution  \eqref{crit-inf}. We solved the FP equation. More precisely, we expressed the Laplace transform of $P_M(t)$ via bi-confluent Heun functions \cite{NIST}. If the initial magnetization is zero, $P_M(t)$ is Gaussian for $t\ll t_c$ and non-Gaussian for $t\gg t_c$; the crossover time is $t_c\sim\sqrt{N}$.

Above or below the critical temperature, the probability distribution $P_M(t)$ obeys Eq.~(\ref{eq:PM-gauss}) and converges on the relaxation time scale $O(\Lambda^{-1})$. The equilibrium distribution is Gaussian near a stable equilibrium. The time evolution of the mean $\langle M\rangle$ obeys Eq.~(\ref{m}). The time evolution of the variance $\Delta^2$ can be derived using an $O(\sqrt{N})$ perturbation to $\langle M\rangle$ and will not be shown explicitly.

For high temperatures ($\beta<\beta_c$), the equilibrium distribution $P_M(\infty)$ is unimodal since there is only one stable equilibrium. For low temperatures ($\beta>\beta_c$), the equilibrium distribution $P_M(\infty)$ is bimodal with peaks of width $\propto \sqrt{N}$ around $M_\pm=\frac{1}{2}\,N\,(1\pm m_\infty)$. Beyond the relaxation time scale $O(\Lambda^{-1})$, the system spends almost all time in the proximity of the peaks, yet occasionally it leaves one peak and reaches the other. The transition time scales exponentially, $e^{CN}$, with system size $N$. The amplitude $C=C(\beta)$ diverges as $\beta\to\infty$ and vanishes as $\beta\to \beta_c$; see Section \ref{sec:ed-tt} for calculation of $C(\beta)$ in general mean-field IMs.

\subsection{Nonzero external magnetic field}
\label{sec:im-non}

Now we impose an external magnetic field $h$ on the IM. We assume $h>0$ since this case is equivalent to the $h<0$ case by flipping the signs of $s_i$ and $h$ simultaneously in the Hamiltonian. The spin-flip rate is
\begin{equation}
    w_i=\frac{1-s_i\tanh[\beta(h+m)]}{2}\,. 
\end{equation}
The magnetization satisfies 
\begin{equation}
\label{eq:dmdt_IM_non}
    \frac{dm}{dt}=-m+\tanh[\beta(h+m)]\equiv f(\beta,m)
\end{equation}
and the equilibria are given by
\begin{equation}
\label{eq:equi-IM-non}
    \text{arctanh}(m)=\beta(h+m).
\end{equation}
There are no phase transitions for sufficiently strong field, $h\geq1$. When $0<h<1$, there is a phase transition at $\beta_c(h)$. The function $\beta_c(h)$ is given implicitly by Eq.~(\ref{eq:equi-IM-non}) and its derivative in $m$, i.e., $1/(1 - m^2)=\beta$, or
\begin{equation}
\label{eq:beta_c_m_c}
    m=-\sqrt{1-\beta^{-1}},
\end{equation}
which together with \eqref{eq:equi-IM-non} yield
\begin{equation}
\label{eq:beta_h}
    h=\sqrt{1-\beta^{-1}}-\beta^{-1}\text{arctanh}\left(\sqrt{1-\beta^{-1}}\right).
\end{equation}

A straightforward asymptotic analysis of \eqref{eq:beta_h} gives
\begin{subequations}
\begin{align}
&\beta_c(h)\to1+\left(\frac{3h}{2}\right)^{2/3}\qquad\text{as}\quad h\to0^+ ,\\
&\beta_c(h)\to-\frac{\log(2(1-h))}{2(1-h)}~\quad\text{as}\quad h\to1^-.
\end{align}
\end{subequations}

\begin{figure}
\centering
    \includegraphics[width=0.46\textwidth]{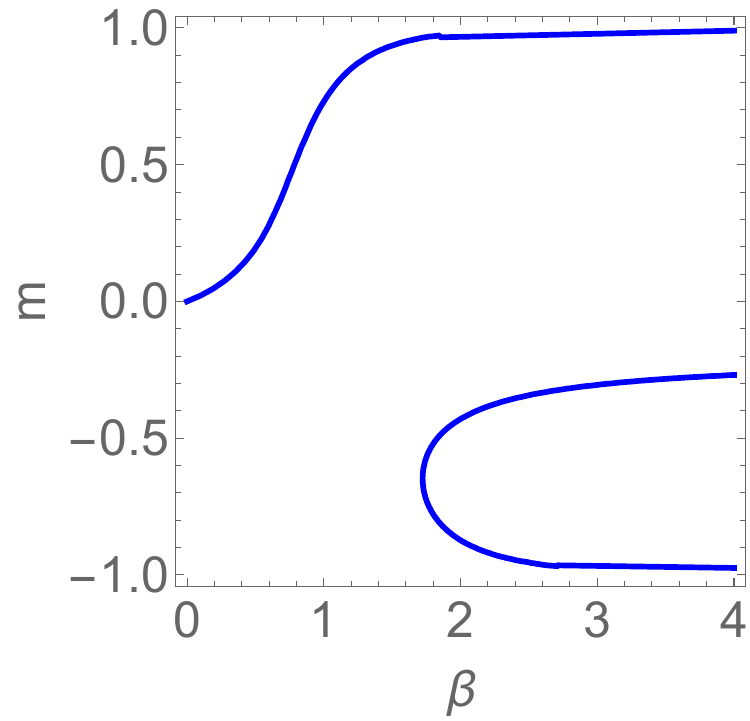}
\caption{The equilibrium magnetization $m_+, m_0, m_-$, top to bottom, versus $\beta=T^{-1}$ when $h=\frac{1}{5}$. The positive equilibrium $m_+(\beta)>0$ exists in the entire temperature range. Two negative equilibria are created in a saddle-node bifurcation at $\beta_c(h)$ implicitly determined by Eq.~\eqref{eq:beta_h}. At the bifurcation point, $m_c(h)=m_0(\beta_c)=m_-(\beta_c)=-\sqrt{1-\beta_c^{-1}}$. In the present case, $\beta_c(\frac{1}{5})\approx 1.72286$ and $m_c(\frac{1}{5})\approx -0.647742$. In the zero-temperature limit, $m_\pm(\infty) = \pm 1$ and $m_0(\infty)=-\frac{1}{5}$; generally,
$m_0(\infty)=-h$ when $0<h<1$.}
\label{Fig:mmm}
\end{figure}

Hereinafter, we consider the range $0<h<1$ of the magnetic field where the phase transition can occur. The positive equilibrium $m_+>0$ persists for all $\beta$. Two negative equilibria $0>m_0>m_-$ are created in a saddle-node bifurcation as $\beta$ increases past $\beta_c(h)$, i.e., as the temperature drops below the critical temperature. As an illustration, see Fig.~\ref{Fig:mmm}. 

Near the bifurcation point $(\beta,m)=(\beta_c,m_c)$, Eq.~(\ref{eq:dmdt_IM_non}) acquires the saddle-node normal form
\begin{equation}
\label{m-bh}
    \frac{dm}{dt}=\frac{\partial f}{\partial\beta}(\beta-\beta_c)+\frac{1}{2}\frac{\partial^2f}{\partial m^2}(m-m_c)^2
\end{equation}
in the leading order. The derivatives in \eqref{m-bh} are computed at $(\beta,m)=(\beta_c,m_c)$ and $\frac{\partial f}{\partial m}(\beta_c,m_c)=0$ is used. 

At criticality, Eq.~\eqref{m-bh} simplifies to
\begin{equation}
\label{m-h-crit}
\frac{dm}{dt} = \sqrt{\beta_c(\beta_c-1)}\, (m-m_c)^2.
\end{equation}
In computing the derivative $\frac{\partial^2f}{\partial m^2}$, we have used the definition of $f$, see \eqref{eq:dmdt_IM_non}, and $m_c=-\sqrt{1-\beta_c^{-1}}$, cf. \eqref{eq:beta_c_m_c}. If the initial magnetization is slightly below the critical, $0<m_c-m\ll 1$, we can rely on \eqref{m-h-crit} leading to 
\begin{equation}
\label{z_c}
m_c-m\simeq  [\beta_c(\beta_c-1)]^{-\frac{1}{2}} t^{-1}
\end{equation}
Thus, the critical dynamic exponent is $z_c=1$. When $m>m_c$, the quantity $m-m_c$ increases with time and the expansion near the bifurcation point becomes invalid. So, we cannot use \eqref{m-h-crit}, but the final state is obvious: $m(t=\infty)=m_+(\beta_c)$. 

In general, near any equilibrium $m_*=m_+$, $m_0$, or $m_-$, we have $m-m_*\sim e^{-\Lambda(m_*)t}$, where the inverse relaxation time is, using Eq.~(\ref{eq:equi-IM-non}),
\begin{equation}
\label{Lambda}
    \Lambda(m_*)\equiv-\frac{\partial f}{\partial m}(\beta,m_*)=1-\beta(1-m_*^2).
\end{equation}

Recall, that $m_+$ is defined for all $\beta$, and as $\beta$ increases from 0 to $\infty$, the equilibrium solution $m_+(\beta)$ increases from 0 to $1$, see Fig.~\ref{Fig:mmm}. The limiting behaviors 
\begin{equation}
    m_+\to\begin{cases}
        \beta h, & \beta\to0^+\\
        1-2e^{-2\beta(h+1)}, & \beta\to+\infty
    \end{cases}
\end{equation}
follow from \eqref{eq:equi-IM-non}. Comparing these limiting behaviors with \eqref{Lambda} we see that $\Lambda(m_+)\approx 1$ in the extremal cases. Generally, a more laborious analysis of \eqref{Lambda} reveals that the inverse relaxation time $\Lambda(m_+)=O(1)$ for all $\beta$.

The equilibrium solution $m_0(\beta)$ is defined for $\beta\geq \beta_c$ and increases from $m_c$ to $-h$ as $\beta$ increases from $\beta_c$ to $\infty$; see Fig.~\ref{Fig:mmm} for illustration in the case when $h=\frac{1}{5}$. A straightforward asymptotic analysis gives more precise limiting behaviors
\begin{align*}
    m_0\to
    \begin{cases}
        m_c+\sqrt{-2\frac{\frac{\partial f}{\partial\beta}(\beta_c,m_c)}{\frac{\partial^2f}{\partial m^2}(\beta_c,m_c)}(\beta-\beta_c)}, & \beta\to\beta_c^+\\
        -h-\text{arctanh}(h)\beta^{-1}, & \beta\to+\infty
    \end{cases}
\end{align*}

The equilibrium solution $m_-(\beta)$ is also defined in the range $\beta\geq \beta_c$. One observes that $m_-(\beta)$ decreases from $m_c$ to $-1$ as $\beta$ increases from $\beta_c$ to infinity. The limiting behaviors of $m_-$ are 
\begin{align*}
    m_-\to\begin{cases}
        m_c-\sqrt{-2\frac{\frac{\partial f}{\partial\beta}(\beta_c,m_c)}{\frac{\partial^2f}{\partial m^2}(\beta_c,m_c)}(\beta-\beta_c)}, & \beta\to\beta_c^+\\
        -1+2e^{2\beta(h-1)}, & \beta\to+\infty
    \end{cases}.
\end{align*}
Thus, $\Lambda(m_0)$ is $O(\sqrt{\beta-\beta_c})$ as $\beta\to\beta_c^+$, which is a slow growth, $O(\beta)$ as $\beta\to\infty$, which is a fast growth, and $O(1)$ otherwise, while $\Lambda(m_-)$ is $O(\sqrt{\beta-\beta_c})$ as $\beta\to\beta_c^+$, which is a slow decay, and $O(1)$ otherwise.
By contrast, the inverse relaxation time $\Lambda$ in Eq.~(\ref{m-below}) for zero external magnetic field is $O(\beta-\beta_c)$ as $\beta\to\beta_c^+$.

\subsection{Equilibrium distribution and transition time}
\label{sec:ed-tt}

Now we extend our analysis beyond the classical IM and consider general mean-field IMs whose Hamiltonian ${\cal H}$ is an arbitrary function of $m$. Our goal is to derive the equilibrium distribution and to use this distribution to calculate transition times between two stable equilibria.

To solve the master equation (\ref{MasterEq}), we can rewrite it in matrix form $\frac{dP}{dt}=AP$ where $P\equiv[P_0,\ldots,P_N]^T$ and $A$ is an $(N+1)\times(N+1)$ tridiagonal matrix with
\begin{align*}
    A_{i,i+1}=W_-(i),\quad &i=1,\cdots,N;\\
    A_{i,i-1}=W_+(i-2),\quad &i=2,\cdots,N+1;\\
    A_{i,i}=-W_-(i-1)-W_+(i-1),\quad &i=1,\cdots,N+1,
\end{align*}
where we define $W_-(0)=W_+(N)=0$. The spectrum of the matrix $A$ has a simple zero eigenvalue, while all other eigenvalues are strictly negative. Thus, any initial condition converges onto the null eigenvector of $A$, which is the probability density function (PDF) at thermodynamic equilibrium $P_\text{eq}$. The total probability at any time must satisfy the normalization condition $\sum_MP_M=1$, and the time evolution conserves the total probability.

To determine the equilibrium PDF, we use the cumulative distribution function $Q_M(t)=\sum_{I=0}^{M-1}P_I(t)$, where $M=1,\ldots,N$ with $Q_{N+1}=1$.  Equation (\ref{MasterEq}) becomes
\begin{equation}
    \frac{dQ_M}{dt}=-W_+(M-1)P_{M-1}+W_-(M)P_M.
\end{equation}
Setting $\frac{dQ_M}{dt}=0$ yields the recurrence for $P_M$ which we solve to find
\begin{equation}
\label{eq:rec-sol}
    -\log(P_M)=-\log(P_0)+\sum_{I=1}^M\log\left(\frac{W_-(I)}{W_+(I-1)}\right)
\end{equation}
The constant $P_0$ is determined from the normalization condition. Note that the master equation (\ref{MasterEq}) can be interpreted as a birth-death process \cite{karlin1957classification}, and the equilibrium distribution (\ref{eq:rec-sol}) was originally derived in this branch of population dynamics.

When $N\to\infty$, it is convenient to transform  $I$ and $M$ into the variables $u=\frac{2I}{N}-1$ and $m=\frac{2M}{N}-1$ which remain of the order of unity. We also rescale the rates 
\begin{align}
\begin{split}
    w_-(u)&\equiv\epsilon W_-(I)=\frac{1}{4}(1+u)(1-\tanh[-\beta{\cal H}'(u)]),\\
    w_+(u)&\equiv\epsilon W_+(I)=\frac{1}{4}(1-u)(1+\tanh[-\beta{\cal H}'(u)]),
\end{split}\label{eq:wpmu}
\end{align}
with $\epsilon=\frac{1}{N}$. Replacing summation by integration we recast \eqref{eq:rec-sol} into
\begin{align}
    \begin{split}
    & -\epsilon\log(\rho_\text{eq}(m))+\epsilon\log(\rho_\text{eq}(-1))\\
    & = \int_{-1}^m\frac{1}{2}\log\left(\frac{w_-(u)}{w_+(u)}\right)du+O(\epsilon)\\
    & \equiv  U(m)-U(-1)+O(\epsilon)
    \end{split}
    \label{eq:MEeq}
\end{align} 
for the equilibrium distribution $\rho(m)\equiv P_M$.  Using Eq.~(\ref{eq:wpmu}), we further simplify \eqref{eq:MEeq}
\begin{equation}
\label{eq:Um}
\begin{split}
&  U(m) = \beta{\cal H}(m)+V(m)\\
& V(m)\equiv\int^m du\,\text{arctanh}(u).
\end{split}
\end{equation}
Thermodynamically, $U(m)/\beta$ is a free energy.

The equilibrium distribution $\rho_\text{eq}(m)$ peaks at each stable equilibrium $m_*$. To find the peak width $\sigma$, we plug $U(m)\to U(m_*)+\frac{1}{2}U''(m_*)(m-m_*)^2$ into Eq.~(\ref{eq:MEeq}) and obtain $\sigma=O\left(\sqrt{\epsilon/U''(m_*)}\right)$, so the widths of different peaks are on the same order in $\epsilon$. Meanwhile, the peak height $\rho_\text{eq}(m_*)$ is beyond all algebraic orders in $\epsilon$ and thus depends sensitively on $U(m_*)$. Thus, to compare the areas of different peaks, it suffices to compare the peak heights and disregard the peak widths.

If there are two stable equilibria $m_\pm$, then we denote the transition probability from $m_-$ to $m_+$ by $\rho_{-+}$ and the transition probability from $m_+$ to $m_-$ by $\rho_{+-}$. The transition time, i.e., the inverse transition probability, is the time spent in the origin phase, which must be proportional to the equilibrium probability, i.e., the peak area, of the origin phase. This relation yields, to leading order,
\begin{equation}
    \rho_\text{eq}(m_-)\rho_{-+}=\rho_\text{eq}(m_+)\rho_{+-}.
\end{equation}
The information on the two peaks enables us to calculate the ratio $\rho_{-+}/\rho_{+-}$, but not $\rho_{-+}$ and $\rho_{+-}$ individually.

The missing information is that the transition probability $\rho_{-+}$ or $\rho_{+-}$ is dominated by the probability to reach the least probable state between $m_-$ and $m_+$, i.e., the unstable equilibrium (saddle) $m_0$. Once $m_0$ is reached, the transition happens with an $O(1)$ probability. Due to the Markov property of the birth-death process, which is a continuous-time Markov chain, the transition probability from $m_-$ or $m_+$ to $m_0$ is the ratio between the equilibrium probabilities of these two states, so
\begin{equation}\label{eq:MEtr}
    \rho_\text{eq}(m_-)\rho_{-+} = \rho_\text{eq}(m_+)\rho_{+-}=\rho_\text{eq}(m_0).
\end{equation}
Denoting $C_{-+}\equiv-\epsilon\log(\rho_{-+})$ and $C_{+-}\equiv-\epsilon\log(\rho_{+-})$, Eqs.~(\ref{eq:MEeq}) \& (\ref{eq:MEtr}) yield
\begin{equation}\label{eq:Cmppm}
    C_{-+}=U(m_0)-U(m_-), ~C_{+-}=U(m_0)-U(m_+).
\end{equation}
Such relations are perhaps best known in transition state theory for chemical reactions \cite{truhlar1996current}, which states that the reaction rate depends on the activation energy.

\section{Phase diagrams and dynamics of feedback Ising model} 
\label{FIM1}

In the FIM defined by the Hamiltonian (\ref{ham-F}), the coupling $\gamma m+1$ is a linear function of magnetization. Specifically, for $0<\gamma\leq1$, the coupling is positive for all $m\in(-1,1)$, so the interaction is totally ferromagnetic. However, for $\gamma>1$, the coupling becomes negative for $m\in(-1,-1/\gamma)$, indicating that the interaction is partly antiferromagnetic. Furthermore, the FIM with $\gamma>0$ involves both 2-spin and 3-spin interactions.

For the FIM on any simple graph, the spin-flip rate for the Glauber dynamics can be written \cite{KRB} as 
\begin{equation}
\label{eq:w_i}
    w_i=\frac{1}{2}\left[1-s_i\tanh\left(-\beta\frac{\Delta{\cal H}_i}{2}\right)\right],
\end{equation}
where $\Delta{\cal H}_i\equiv{\cal H}_\text{FIM}(s_i=+1)-{\cal H}_\text{FIM}(s_i=-1)$ is the Hamiltonian change due to flipping $s_i$. Using Eq.~(\ref{ham-F}) and noting that $m$ changes by $2/N$ we get 
\begin{equation}
\label{delta-H}
    \frac{\Delta{\cal H}_i}{2}=-h-\frac{1}{N}(1+\gamma m)\sum_{j\in\langle i\rangle}s_j-\frac{1}{N^2}\gamma\sum_{\langle i,j\rangle}s_is_j.
\end{equation}
On the complete graph, i.e., for the Curie-Weiss model, \eqref{delta-H} simplifies to
\begin{equation}\label{eq:gm3o2}
    \frac{\Delta{\cal H}_i}{2}=-h-m\left(1+\frac{3}{2}\gamma m\right)\equiv g(m)
\end{equation}
in the $N\rightarrow\infty$ limit. Alternatively, this can be derived using $g(m)={\cal H}'(m)$ and simplifying the Hamiltonian (\ref{ham-F})  to  ${\cal H}(m)=-hm-\frac{1}{2}(1+\gamma m)m^2$ in the $N\rightarrow\infty$ limit. Using the general relation (\ref{si}) with spin-flip rate (\ref{eq:w_i}) and summing over $i$ we arrive at the evolution equation
\begin{equation}
\label{eq:dmdt-general}
    \frac{dm}{dt}=-m+\tanh[-\beta g(m)].
\end{equation}

In this section, we first construct phase diagrams and their associated bifurcation diagrams in the FIM with or without an external magnetic field. Then, we analyze dynamics of discrete probability distributions in the FIM, including convergence to a stable equilibrium and transition between two stable equilibria. We use the temperature $T\equiv\beta^{-1}$ rather than the inverse temperature $\beta$ in the diagrams such that the low temperature regime forms a compact interval in $T$; we use $T$ and $\beta$ interchangeably in the calculations.

\begin{figure}
\centering
 \includegraphics[width=0.44\textwidth]{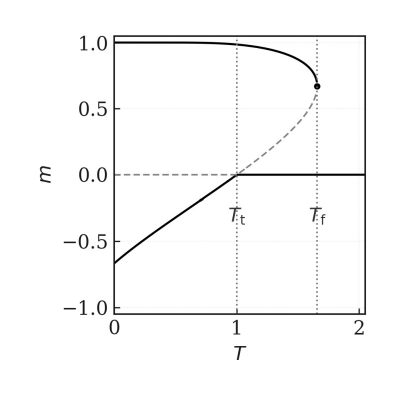}
\caption{The equilibrium magnetization $m$ versus the temperature $T$ in the FIM with  $h=0$ and $\gamma=1$. The magnetization is implicitly given by Eq.~\eqref{m-F:SS} with $\gamma=1$. Solid curves: stable branches. Dashed curves: unstable branches. The stable values of the magnetization are $m=1$ and $m=-\frac{2}{3}$ when $T=0$.}
\label{Fig:beta-m}
\end{figure}

\subsection{Zero external magnetic field}
\label{sec:FIM-zero}

When $h=0$, the magnetization satisfies
\begin{equation}
\label{m-F}
\frac{dm}{dt}= -m + \tanh\!\Bigl[\beta\,m\bigl(1+\tfrac{3}{2}\gamma m\bigr)\Bigr]\equiv f(\beta,m)
\end{equation}
and the equilibrium is determined by
\begin{equation}
\label{m-F:SS}
m = \tanh\!\Bigl[\beta\,m\bigl(1+\tfrac{3}{2}\gamma m\bigr)\Bigr].
\end{equation}

In addition to the disordered phase $M_0$ with magnetization $m=0$, which is a solution of~\eqref{m-F:SS} for all temperatures, there are two ordered phases $M_\pm$ with magnetizations $m_\pm$ that exist for sufficiently low temperatures, $\beta\ge \beta_f$. The curves $m_\pm(T)$ for the temperatures $T\leq T_f$ shown in Fig.~\ref{Fig:beta-m} correspond to the FIM with the feedback parameter having the marginal value $\gamma=1$. (Recall that for the FIM, the marginal value separates the parameter range $0<\gamma<1$ where interactions are ferromagnetic from  the parameter range $\gamma>1$ where interactions are partly antiferromagnetic.) 

At $\beta=\beta_f$, the two branches meet in a saddle-node (fold) bifurcation, so $m_+(\beta_f)=m_-(\beta_f)\equiv m_f$.  To find the bifurcation point $(\beta_f,m_f)$, we  invoke the implicit-function theorem: at the  bifurcation point, the derivative $\frac{dm}{d\beta}$ diverges. Differentiating Eq.~\eqref{m-F:SS} with respect to $\beta$ 
\begin{equation}
\frac{dm}{d\beta}
\bigl[1 - \beta(1-m^{2})(1+3\gamma m)\bigr]
   = m(1-m^{2})\bigl(1+\tfrac{3}{2}\gamma m\bigr)
\end{equation}
and setting $\frac{dm}{d\beta}=\infty$ we obtain
\begin{subequations}
\label{beta-m-f}
\begin{equation}
\label{beta-f}
\beta_f^{-1}=(1-m_f^{2})(1+3\gamma m_f) 
\end{equation}
which we combine with \eqref{m-F:SS} to find
\begin{equation}
\label{m-f}
m_f = \tanh\!\left[\frac{m_f\bigl(1+\tfrac32\gamma m_f\bigr)}
                           {(1+3\gamma m_f)(1-m_f^{2})}\right]
\end{equation}
\end{subequations}
If the feedback parameter is marginal ($\gamma=1$), we have $m_f \approx 0.670\,784\,995\,198$ and $\beta_f \approx 0.603\,522\,774\,698$.

The second critical temperature remains $\beta_t = 1$, identical to the Curie-Weiss value. At $\beta=\beta_t$, the $M_-$ branch merges with the disordered state $M_0$ in a transcritical bifurcation. For $\beta>\beta_t$ the disordered solution is unstable, whereas in the intermediate range $\beta_f<\beta<\beta_t$ both $M_0$ and the highly magnetised phase $M_+$ are stable.

Equation (\ref{m-F}) simplifies to
\begin{equation}
    \frac{dm}{dt}=(\beta-1)m+\frac{3}{2}\,\beta\gamma m^2+\ldots
\end{equation}
near the bifurcation point $(\beta,m)=(1,0)$. Therefore, the equilibrium $m=0$ persists as $\beta$ varies. At criticality, $\frac{dm}{dt}\simeq \frac{3}{2}\gamma m^2$, yielding an algebraic decay 
\begin{equation}
m\simeq - \frac{2}{3\gamma}\,t^{-1}
\end{equation}
if the initial magnetization is negative, $m(0)<0$. Thus, the critical dynamic exponent of the transcritical bifurcation is $z_t=1$. This exponent value agrees with that of the fold bifurcation, cf. \eqref{z_c}, so these two bifurcations differ near criticality but agree at criticality. If $m(0)>0$, the magnetization increases until it reaches a positive root of \eqref{m-F:SS} at criticality, $m = \tanh[m(1+\tfrac{3}{2}\gamma m)]$. 

At the transcritical temperature $T_t=1$, the system loses bistability since the equilibrium $m=0$ is marginally stable. However, the system regains bistability as the temperature increases from $T_t$ until $T_f$ where bistability is lost again. In contrast, a temperature increase in the classical IM always causes the system to lose bistability as shown in Sec.~\ref{sec:CW}. Thus, temperature-induced bistability is a unique feature of the FIM.

Using \eqref{beta-m-f} we find the large $\gamma$ behavior of the temperature $T_f$ and magnetization $m_f$
\begin{subequations}
\begin{align}
\label{T-f:large}
T_f(\gamma) &= 3\,m_\infty\bigl(1-m_\infty^{2}\bigr)\gamma + O(1)\\
\label{m-f:large}
m_f(\gamma) &= m_\infty + O(\gamma^{-1})
\end{align}
\end{subequations}
where $m_\infty=m_f(\infty)$ is an implicit solution of [cf. \eqref{m-f}]
\begin{equation*}
m_\infty = \tanh\!\left[\frac{m_\infty}
                           {2(1-m_\infty^{2})}\right]
\end{equation*}
Numerically $m_\infty\approx  0.796388$. In contrast with $T_f(\gamma)$ that varies with $\gamma$ and diverges when $\gamma\to\infty$, the transcritical threshold remains constant: $T_t = 1$.

When the temperature approaches zero, $\beta\to\infty$, the $M_+$ phase approaches to the plus ground state. A more precise asymptotic
\begin{equation}
1-m_+ \simeq 2\,e^{-\beta(2+3\gamma)}
\end{equation}
follows from Eq.~\eqref{m-F:SS}. The magnetization $m_-$ in the $M_-$ phase exhibits a richer behavior:
\begin{equation}\label{eq:m_-_h0}
m_-\simeq 
\begin{cases}
-1+2 e^{-\beta(2-3\gamma)}                                                                      & 0<\gamma < \frac{2}{3}\\
-1 + \frac{\log \beta - \log(\log \beta)}{2\beta}                                              & \gamma = \frac{2}{3}\\
-\frac{2}{3\gamma} + \beta^{-1}\text{arctanh}(\frac{2}{3\gamma}) & \gamma > \frac{2}{3}
\end{cases}
\end{equation}
Thus in the $\beta\to\infty$ limit, the $M_-$ phase approaches to the minus ground state only when $\gamma \leq \frac{2}{3}$. The $M_-$ phase for $\gamma>\frac{2}{3}$ with both up-spins and down-spins is called a mixed phase (MP); see Ref.~\cite{ma2025mixed} for a detailed analysis of such MPs in the general FIM at zero temperature.

\begin{figure*}
    \centering
    \subfigure[]{\includegraphics[width=0.24\textwidth]{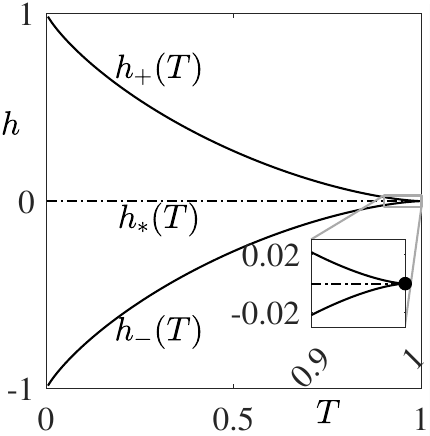}\label{fig:FIM_Th_g0}}
    \subfigure[]{\includegraphics[width=0.24\textwidth]{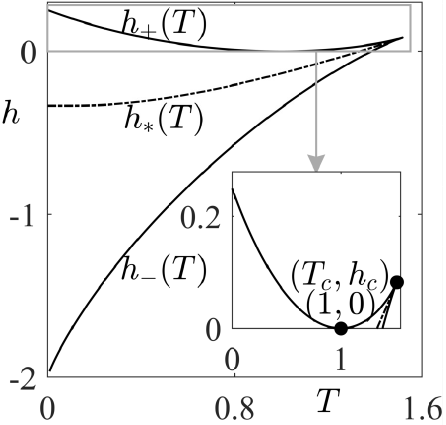}\label{fig:FIM_Th_g23}}
    \subfigure[]{\includegraphics[width=0.24\textwidth]{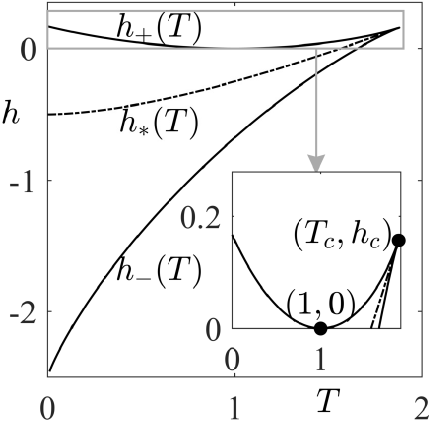}\label{fig:FIM_Th_g1}}
    \subfigure[]{\includegraphics[width=0.24\textwidth]{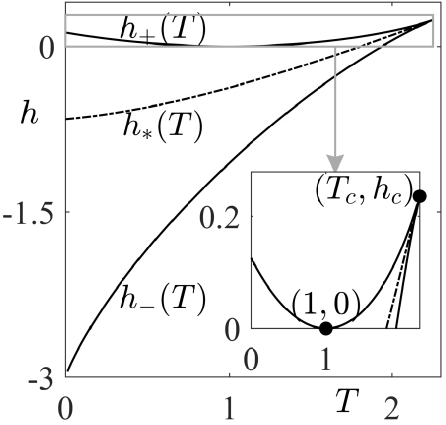}\label{fig:FIM_Th_g43}}
    \caption{Phase diagrams of the FIM on the $(T,h)$-plane for (a) $\gamma=0$; (b) $\gamma=\frac{2}{3}$; (c) $\gamma=1$; 
    and (d) $\gamma=\frac{4}{3}$. There are two stable equilibria $m_\pm$ and one unstable equilibrium $m_0$ satisfying $m_+>m_0>m_-$ 
    in the wedge between $h_\pm(T)$ and one stable equilibrium outside this wedge. The Maxwell curve $h_*(T)$ where $m_\pm$ are equally 
    probable is derived in Sec.~\ref{sec:Maxwell}.}
    \label{fig:FIM_Th}
\end{figure*}

\begin{figure}
    \centering
    \includegraphics[width=0.3\textwidth]{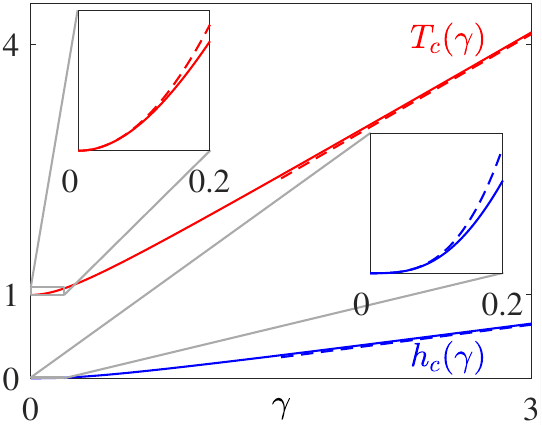}
    \caption{Plot of $T_c$ (red) and $h_c$ (blue) as functions of $\gamma$. The dashed curves show asymptotic behavior 
     predicted by Eqs.~(\ref{eq:Thc_g_0}--\ref{eq:Thc_g_inf}).}
    \label{fig:FIM_Tc_hc}
\end{figure}

\subsection{Nonzero external magnetic field}
\label{sec:FIM-non}

Let us construct the phase diagram of the FIM in the $(T,h)$ half-plane, $T\geq 0$, for fixed $\gamma>0$. We shall also compare with the results for the Curie-Weiss model ($\gamma=0$). 

The magnetization $m$ satisfies the ODE
\begin{equation}
\label{eq:dmdt-FIM-non}
    \frac{dm}{dt}=-m+\tanh\!\left[T^{-1}\left(\frac{3}{2}\gamma m^2+m+h\right)\right].
\end{equation}
Therefore, the equilibria are the solutions of 
\begin{align}
h=T\text{arctanh}(m)-\frac{3}{2}\gamma m^2-m.
\end{align}
The locus of phase transitions is given by $dT/dm=0$ or equivalently $dh/dm=0$, which together yield a parametric curve on the $(T,h)$-plane:
\begin{align}
\begin{split}\label{eq:Tm_hm}
& T(m)=T_0(m)+\gamma T_1(m) \\
& h(m)=h_0(m)+\gamma h_1(m) \\
&T_0(m)\equiv1-m^2, \quad T_1(m)\equiv3mT_0(m)\\
&h_0(m)\equiv(1-m^2)\text{arctanh}(m)-m\\
&h_1(m)\equiv3mh_0(m)+\tfrac{3}{2}m^2.
\end{split}
\end{align}
The magnetization varies in the interval $m\in[-1,1]$ when $\gamma\in[0,\frac{1}{3}]$ and $m\in[-1/(3\gamma),1]$ when $\gamma>\frac{1}{3}$. For any $\gamma\geq0$, the locus of phase transitions on the $(T,h)$-plane consists of an upper branch $h_+(T)$ and a lower branch $h_-(T)$ meeting at a cusp $(T_c,h_c)$; see Fig.~\ref{fig:FIM_Th}. Explicitly, we have $(T_c,h_c)=(T(m_c),h(m_c))$ where 
\begin{equation}
m_c = \frac{-1 + \sqrt{1 + 27 \gamma^2}}{9 \gamma}
\end{equation}
is determined from $T'(m_c)=0$. For any $(T,h)$ in the wedge between $h_\pm(T)$, Eq.~(\ref{eq:dmdt-FIM-non}) has two stable equilibria $m_\pm$ and one unstable equilibrium $m_0$ sandwiched between stable equilibria: $m_+>m_0>m_-$. Outside this wedge, Eq.~(\ref{eq:dmdt-FIM-non}) has one stable equilibrium.

In Fig.~\ref{fig:FIM_Tc_hc}, the location of the cusp $(T_c,h_c)$ is plotted as a function of $\gamma$. Straightforward calculations allow one to derive the asymptotic behaviors
\begin{equation}
\label{eq:Thc_g_0}
    T_c\to1+\frac{9}{4}\gamma^2,\quad h_c\to\frac{9}{8}\gamma^3,
\end{equation}
when $\gamma\to 0$. In the complementary $\gamma\to+\infty$ limit 
\begin{align}
\label{eq:Thc_g_inf}
T_c\to\frac{2}{\sqrt{3}}\gamma+\frac{2}{3}\,, \quad 
 h_c\to\frac{4\alpha-1}{2}\,\gamma+\frac{2\alpha-1}{\sqrt{3}}
\end{align}
where we shortly write $\alpha=\frac{\text{arccoth}\sqrt{3}}{\sqrt{3}}$. 

\begin{figure*}
    \centering
    \subfigure[]{\includegraphics[width=0.3\textwidth]{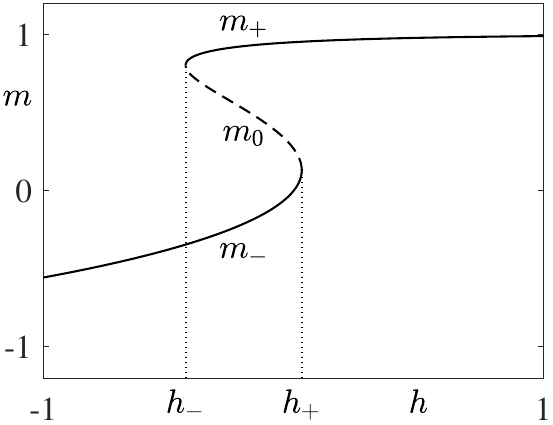}\label{fig:FIM_bif_g43T15}}
    \subfigure[]{\includegraphics[width=0.3\textwidth]{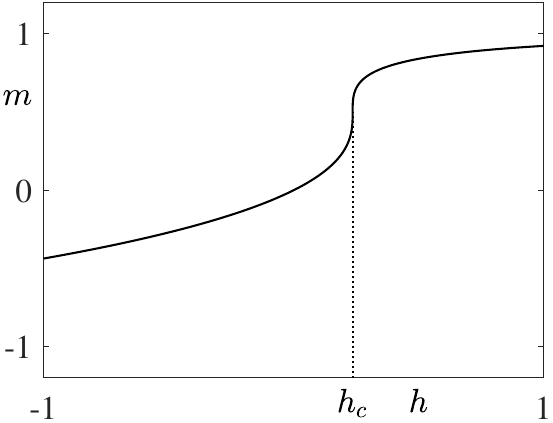}\label{fig:FIM_bif_g43T225}}
    \subfigure[]{\includegraphics[width=0.3\textwidth]{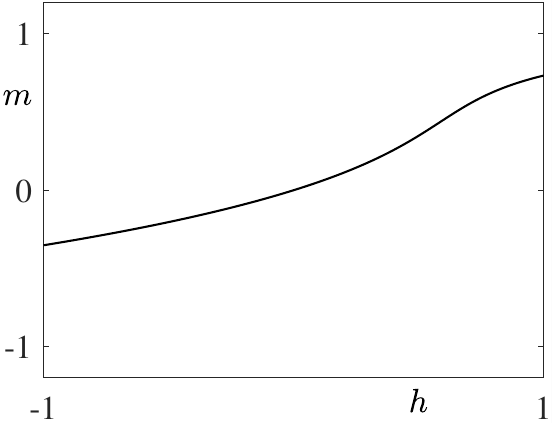}\label{fig:FIM_bif_g43T3}}
    \caption{Bifurcation diagrams of the FIM on the $(h,m)$-plane for $\gamma=\frac{4}{3}$ and (a) $T=1.5$; (b) $T=2.25$; and (c) $T=3$.}
    \label{fig:FIM_bif_h}
\end{figure*}

A vertical slice on the $(T,h)$-plane in Fig.~\ref{fig:FIM_Th} yields a bifurcation diagram on the $(h,m)$-plane for fixed $T$. As shown in Fig.~\ref{fig:FIM_bif_h}, this bifurcation diagram is an S-shaped curve for any fixed $T\in[0,T_c)$ with two fold bifurcations at $h_\pm\equiv h_\pm(T)$, and an increasing curve for any fixed $T\geq T_c$ with a cusp bifurcation at $h=h_c$ when $T=T_c$. In both the classical IM and the FIM, there are two critical magnetic fields when the temperature is fixed below a critical temperature and none above this critical temperature. However, the key difference is that the bifurcation diagram on the $(h,m)$-plane is symmetric under the transformation $(h,m)\rightarrow(-h,-m)$ when $\gamma=0$ and asymmetric when $\gamma>0$.

The contrast between the classical IM and the FIM is more pronounced on the locus of phase transitions. As shown in the insets of Fig.~\ref{fig:FIM_Th}, the upper branch of phase transitions $h_+(T)$ decreases to the cusp $(T_c,h_c)=(1,0)$ in the classical IM but first decreases to a local minimum $(1,0)$ and then increases to the cusp $(T_c,h_c)$ in the FIM; compare Figs.~\ref{fig:FIM_Th_g0} and \ref{fig:FIM_Th_g23}--\ref{fig:FIM_Th_g43}. This feature affects horizontal slices on the $(T,h)$-plane, i.e., bifurcation diagrams on the $(T,m)$-plane for fixed $h$. We can identify the critical values of $h$ where the bifurcation diagram on the $(T,m)$-plane changes qualitatively. The classical IM has three critical values: $h_\pm(0)$ and $h_c=0$. The FIM possesses four critical values: $h_\pm(0)$, $h_c$, and 0. 

Essentially, the FIM features a competition between 2-spin interaction, making the maximum coercive field $h_+(0)$ at zero temperature, and 3-spin interaction, making the maximum coercive field $h_c$ at the cusp. We can identify $\gamma_*=1.0257064\ldots$ such that $h_c\in(0,h_+(0))$ when $\gamma\in(0,\gamma_*)$ and $h_c>h_+(0)$ when $\gamma>\gamma_*$; compare Figs.~\ref{fig:FIM_Th_g23} \& \ref{fig:FIM_Th_g43}. Thus, 2-spin interaction dominates when $\gamma\in(0,\gamma_*)$, while 3-spin interaction dominates when $\gamma>\gamma_*$. Interestingly, $\gamma_*$ only slightly exceeds 1, so the curve $h_+(T)$ is almost symmetric with respect to $(1,0)$ when $\gamma=1$; see Fig.~\ref{fig:FIM_Th_g1}.

\begin{figure*}
    \centering
    \subfigure[]{\includegraphics[width=0.3\textwidth]{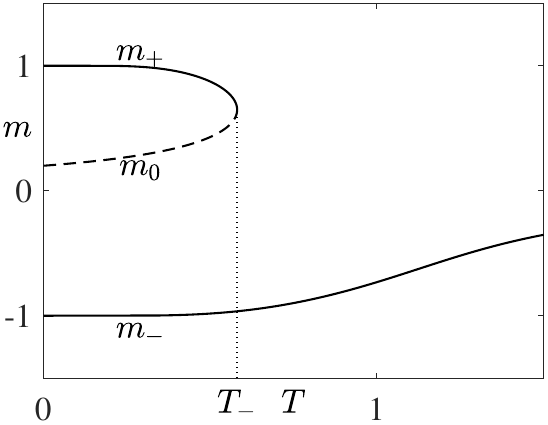}\label{fig:IM_bif_h-02}}
    \subfigure[]{\includegraphics[width=0.3\textwidth]{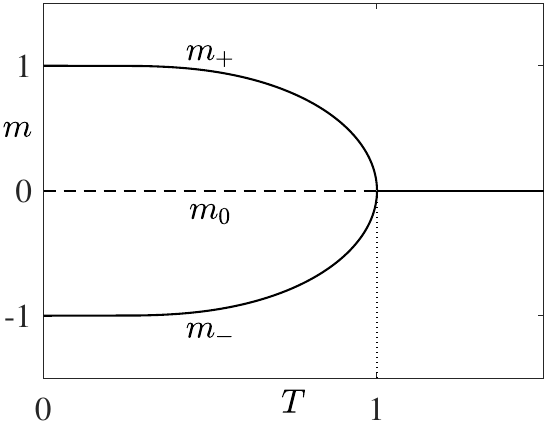}\label{fig:IM_bif_h0}}
    \subfigure[]{\includegraphics[width=0.3\textwidth]{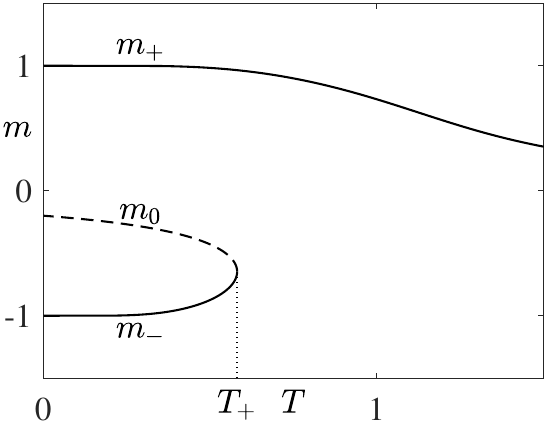}\label{fig:IM_bif_h02}}
    \caption{Bifurcation diagrams of the classical IM on the $(T,m)$-plane for (a) $h=-0.2$; (b) $h=0$; and (c) $h=0.2$. The subfigures (a) and (c) 
    are essentially identical reflecting the invariance of the IM upon the transformation $(s,h)\to (-s,-h)$. The subfigure (c) coincides with 
    Fig.~\ref{Fig:mmm} where we plotted magnetization versus $\beta$ rather than $T$ and additionally provided precise numerical coordinates 
    of the point where the $m_0$ and $m_-$ branches meet.}
    \label{fig:IM_bif_T}
\end{figure*}

As shown in Figs.~\ref{fig:IM_bif_h-02} \& \ref{fig:IM_bif_h02}, in the classical IM, the bifurcation diagram on the $(T,m)$-plane is a flipped C-shaped curve above an increasing curve for any fixed $h\in(h_-(0),0)$ with a fold at $T_-\equiv h_-^{-1}(h)$ and a flipped C-shaped curve below a decreasing curve for any fixed $h\in(0,h_+(0))$ with a fold at $T_+\equiv h_+^{-1}(h)$. These two diagrams are related by up-down symmetry. We analyzed the latter case in Section \ref{sec:im-non} where $T_+$ was denoted as $T_c$. At the transition point $h=0$ between these two diagrams, the only bifurcation is a pitchfork bifurcation at $T=1$ as shown in Fig.~\ref{fig:IM_bif_h0}. We analyzed this case in Section \ref{sec:im-zero}. Overall, there is a single critical temperature when the magnetic field is within a certain interval and none outside this interval.

In the FIM, we can still denote $T_-\equiv h_-^{-1}(h)$, which is single-valued for $h\in(h_-(0),0)$. However, $h_+^{-1}(h)$ can be multi-valued, and we denote by $T_+$ the value in $(0,1)$ for $h\in(0,h_+(0))$. On the $(T,m)$-plane, this fold $T_+$ connects $m_-$ and $m_0$ on a flipped C-shaped curve as in the classical IM. For $h\in(0,h_c)$, the set $h_+^{-1}(h)$ contains two more values $T_1$ and $T_2$ with $T_2>T_1>1$. On the $(T,m)$-plane, this implies a flipped S-shaped curve with two folds $T_1$ and $T_2$. 
%As $\gamma\to\infty$, i.e., in the 3-spin Curie-Weiss model, this flipped S-shaped curve alone forms the entire bifurcation diagram on the $(T,m)$-plane for $\hat{h}\in(0,\hat{h}_c)$. 
The unique feature of the FIM is that the flipped C-shaped and S-shaped curves can coexist, yielding one, two, or three critical temperatures when the magnetic field is small and positive.

\begin{figure*}
    \centering
    \subfigure[]{\includegraphics[width=0.3\textwidth]{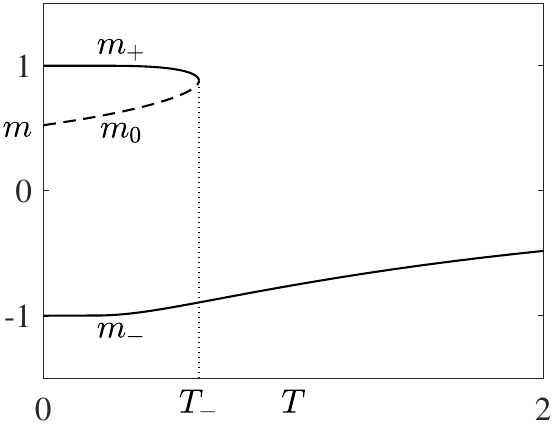}\label{fig:FIM_bif_g23h-08}}
    \subfigure[]{\includegraphics[width=0.3\textwidth]{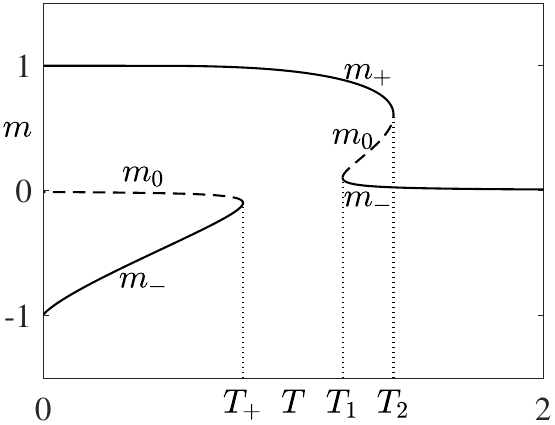}\label{fig:FIM_bif_g23h001}}
    \subfigure[]{\includegraphics[width=0.3\textwidth]{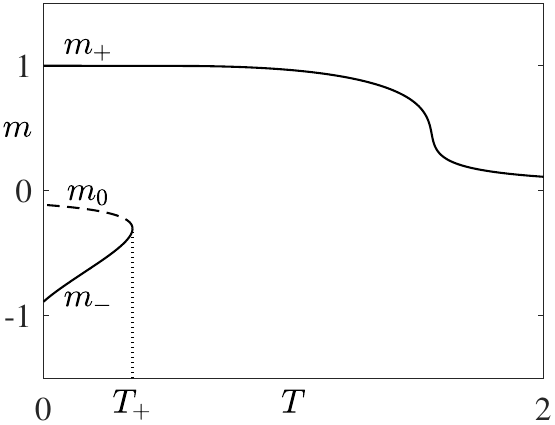}\label{fig:FIM_bif_g23h01}}
    \caption{Bifurcation diagrams of the FIM on the $(T,m)$-plane for $\gamma=\frac{2}{3}$ and (a) $h=-0.8$; (b) $h=0.01$; and (c) $h=0.1$.}
    \label{fig:FIM_bif_T_g23}
\end{figure*}

\begin{figure*}
    \centering
    \subfigure[]{\includegraphics[width=0.3\textwidth]{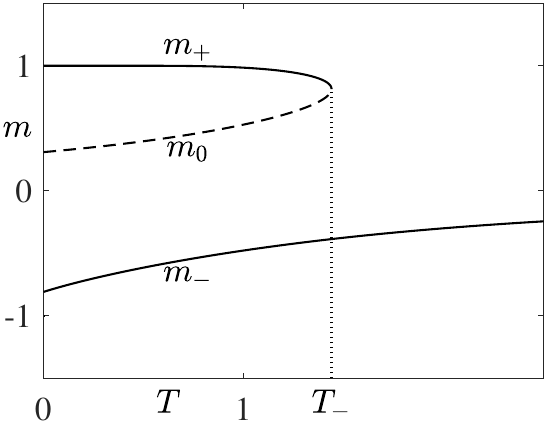}\label{fig:FIM_bif_g43h-05}}
    \subfigure[]{\includegraphics[width=0.3\textwidth]{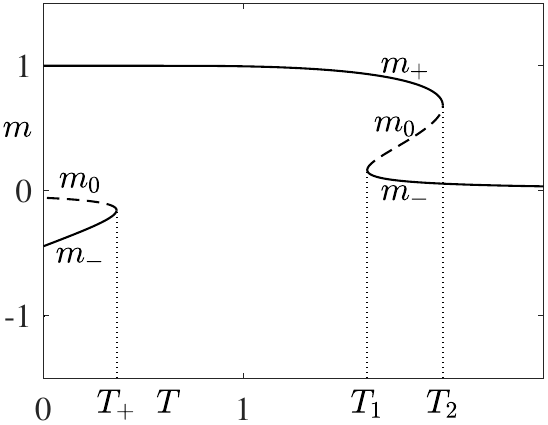}\label{fig:FIM_bif_g43h005}}
    \subfigure[]{\includegraphics[width=0.3\textwidth]{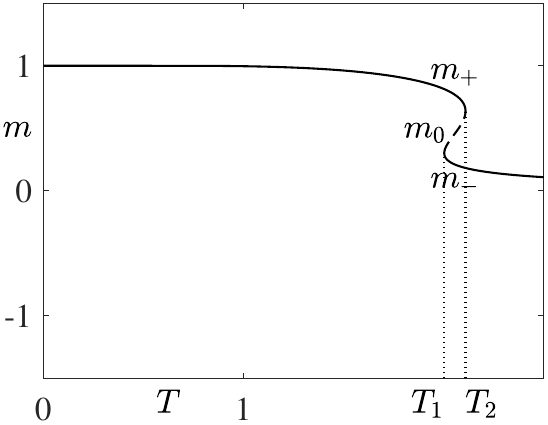}\label{fig:FIM_bif_g43h014}}
    \caption{Bifurcation diagrams of the FIM on the $(T,m)$-plane for $\gamma=\frac{4}{3}$ and (a) $h=-0.5$; (b) $h=0.05$; and (c) $h=0.14$.}
    \label{fig:FIM_bif_T_g43}
\end{figure*}

The sequence of bifurcation diagrams on the $(T,m)$-plane as $h$ increases are shown in Figs.~\ref{fig:FIM_bif_T_g23} \& \ref{fig:FIM_bif_T_g43} respectively for $\gamma\in(0,\gamma_*)$ and $\gamma>\gamma_*$. In both cases, panel (a) shows a one-fold bifurcation diagram for $h\in(h_-(0),0)$, while panel (b) shows a three-fold bifurcation diagram for $h\in(0,\min(h_+(0),h_c))$ exhibiting coexistence between flipped C-shaped and S-shaped curves. As $h$ increases, panel (a) transitions into panel (b) via a non-generic bifurcation diagram at $h=0$ like Fig.~\ref{Fig:beta-m}. Specifically, the fold $T_-$ transitions smoothly to $T_2$ via $T_f$, while the $m_-$ and $m_0$ branches reconnect into two mixed branches with folds $T_+$ and $T_1$ via the transcritical bifurcation at $T_t=1$. Technically, this type of branch reconnection is known as a universal unfolding of the transcritical bifurcation \cite{wiggins}. The number of folds decreases as $h$ increases further. For $\gamma\in(0,\gamma_*)$, the flipped S-shaped curve becomes monotonic as shown in Fig.~\ref{fig:FIM_bif_g23h01} where $h\in(h_c,h_+(0))$. For $\gamma>\gamma_*$, the flipped C-shaped curve disappears as shown in Fig.~\ref{fig:FIM_bif_g43h014} where $h\in(h_+(0),h_c)$. For $h\in(0,h_c)$, the flipped S-shaped curve with the two folds $T_1<T_2$ represents the generic form of temperature-induced bistability where the system first gains and then loses bistability as the temperature increases.

The above analysis reveals discontinuous phase transitions occurring when $T>0$. When $T=0$ and $\gamma>\frac{1}{3}$, there is additionally a continuous (second-order) phase transition at $h_\text{sec}=1-\frac{3}{2}\gamma$ such that $m_-=-1$ when $h\leq h_\text{sec}$ and $|m_-|<1$ when $h>h_\text{sec}$. The latter extends the MP at $h=0$ in Eq.~(\ref{eq:m_-_h0}) to $h\neq0$. In Ref.~\cite{ma2025mixed}, such MPs are shown to be super-stable, i.e., perturbation to the magnetization decays linearly and vanishes in finite time. In Sec.~\ref{sec:conv-mp}, we show that super-stability affects convergence of not only the magnetization but also the probability distribution.

\subsection{Maxwell curve}
\label{sec:Maxwell}

The key dynamical feature of the bistable regime in general mean-field IMs is long-time transitions between the two stable equilibria as analyzed in Section \ref{sec:ed-tt}. In the FIM, there are two stable equilibria $m_\pm$ when $T\in[0,T_c)$ and $h\in(h_-(T),h_+(T))$ as shown in Fig.~\ref{fig:FIM_Th}. Since $m_\pm$ is the more probable state as $h\to h_\pm(T)$, there exists a Maxwell curve $h_*(T)$ where $m_\pm$ are equally probable, or equivalently the two transition probabilities $\rho_{\pm,\mp}$ in Eq.~(\ref{eq:MEtr}) are equal. 

Thermodynamically, $m_\pm$ have equal free energies on the Maxwell curve. Using Eqs.~(\ref{eq:MEeq}--\ref{eq:Um}), $h_*(T)$ is given implicitly by
\begin{equation}
    {\cal M}\equiv T^{-1}{\cal M}_0+{\cal M}_1=0,
\end{equation}
where
\begin{equation}
    {\cal M}_0\equiv{\cal H}(m_+)-{\cal H}(m_-),\quad
    {\cal M}_1\equiv V(m_+)-V(m_-).
\end{equation}
The energy per spin is 
\begin{equation}
\label{energy:FIM}
\mathcal{H}(m) =-\frac{1}{2}(1+\gamma m)m^2-hm
\end{equation}
for the FIM, with `potential energy' 
\begin{equation}
\label{Vm}
\begin{split}
    V(m)                  &=\int_0^m du\, \text{arctanh}(u)\\
    &=m \cdot \text{arctanh}(m) + \frac{1}{2} \log(1 - m^2)\\
    &=\frac{(1+m)\log(1+m) + (1-m)\log(1-m)}{2}
\end{split}
\end{equation}
when $|m|<1$, and $V(\pm 1)=\log(2)$. For each $T\in(0,T_c)$ and $h\in(h_-(T),h_+(T))$, ${\cal M}$ can be evaluated numerically, and the Maxwell curve $h_*(T)$ as the zero level set of the function ${\cal M}(T,h)$ is plotted in the $(T,h)$-plane in Fig.~\ref{fig:FIM_Th}. The upper equilibrium $m_+$ is the more probable state when $h>h_*(T)$, while the lower equilibrium $m_-$ is the more probable state when $h<h_*(T)$. Thus, increasing the magnetic field always favors the upper equilibrium in both the classical IM and the FIM.

Although a unique Maxwell magnetic field exists at a fixed temperature, a unique Maxwell temperature may not exist at a fixed magnetic field. In the classical IM with $\gamma=0$, we have $h_*(T)=0$ as shown in Fig.~\ref{fig:FIM_Th_g0}, so there are infinitely many Maxwell temperatures at $h=0$ and none otherwise. Thus, increasing the temperature favors neither the upper nor the lower equilibrium. However, in the FIM with $\gamma>0$, $h_*(T)$ increases to the cusp $(T_c,h_c)$ as shown in Figs.~\ref{fig:FIM_Th_g23}--\ref{fig:FIM_Th_g43}, so there is a unique Maxwell temperature for any fixed $h\in(h_*(0),h_c)$ and none otherwise. For any fixed $h\in(0,h_c)$, the Maxwell temperature exists in the finite-temperature interval disconnected from $T=0$. The unique feature of the FIM is that the Maxwell temperature is well defined when $h$ is small and either positive or negative, and increasing the temperature favors the lower equilibrium.

\subsection{Transition time between two stable equilibria}
\label{sec:tran-fim}

The Maxwell curve shows where the transition times between the two stable equilibria $m_\pm$ are equal, but the calculation of the two transition times, $O\left(e^{C_{-+}N}\right)$ from $m_-$ to $m_+$ and $O\left(e^{C_{+-}N}\right)$ from $m_+$ to $m_-$, requires the free energy of the saddle $m_0$ as shown in Section \ref{sec:ed-tt}. For any $(T,h)$ in the bistable region
\[
{\cal B}:=\{(T,h):T\in(0,T_c),h\in(h_-(T),h_+(T))\},
\]
the transition rates $C_{-+}$ and $C_{+-}$ predicted by Eq.~(\ref{eq:Cmppm}) \& (\ref{eq:Um}) are explicitly
\begin{align}
\begin{split}
    &C_{-+}=\beta({\cal H}(m_0)-{\cal H}(m_-))+V(m_0)-V(m_-),\\
    &C_{+-}=\beta({\cal H}(m_0)-{\cal H}(m_+))+V(m_0)-V(m_+).
\end{split}
\end{align}
Thus, the transition rate depends on the distance between the origin phase and the saddle. Near the lower boundary $h_-(T)$ of ${\cal B}$, we have $0<m_+-m_0\ll1$, so $0<C_{+-}\ll1$. Similarly, near the upper boundary $h_+(T)$ of ${\cal B}$, we have $0<m_0-m_-\ll1$, so $0<C_{-+}\ll1$. These two cases feature fast transitions near critical points explored in Sect.~\ref{LTR}.

Away from the upper or lower boundary of ${\cal B}$, $m_\pm-m_0$ is always $O(1)$. Near the left boundary $T=0$ of ${\cal B}$, we have $\beta\gg1$, so both $C_{-+}$ and $C_{+-}$ are $O(\beta)$. This case features slow transitions near zero temperature. For a generic $(T,h)\in{\cal B}$ away from all three boundaries of ${\cal B}$, both $C_{-+}$ and $C_{+-}$ are $O(1)$.

Although the above description applies at any $\gamma$, the key difference between the classical IM and the FIM is that the former exhibits the up-down symmetry $C_{-+}(T,h)=C_{+-}(T,-h)$ while the latter does not. This is why the Maxwell curve $h_*(T)$ is $h=0$ in the classical IM but becomes nontrivial in the FIM as shown in Fig.~\ref{fig:FIM_Th}.

\subsection{Convergence to stable equilibrium}\label{sec:conv-mp}

From the analysis in Sections \ref{sec:im-zero} and \ref{sec:im-non}, we can postulate a central dogma for convergence of the probability distribution near a stable equilibrium in general mean-field IMs including the FIM: the probability distribution $P_M(t)$ is Gaussian for all $t$ if and only if the convergence to the stable equilibrium is exponential.

At a finite temperature, algebraic convergence to a stable equilibrium implies proximity to a bifurcation point. In this case, the probability distribution becomes continuous and non-Gaussian on certain time intervals as analyzed in Section \ref{LTR} and Appendix \ref{sec:conv-eq-zero}. At zero temperature, algebraic convergence only occurs when the stable equilibrium is a super-stable MP. 

\begin{figure}
 \centering
    \includegraphics[width=0.46\textwidth]{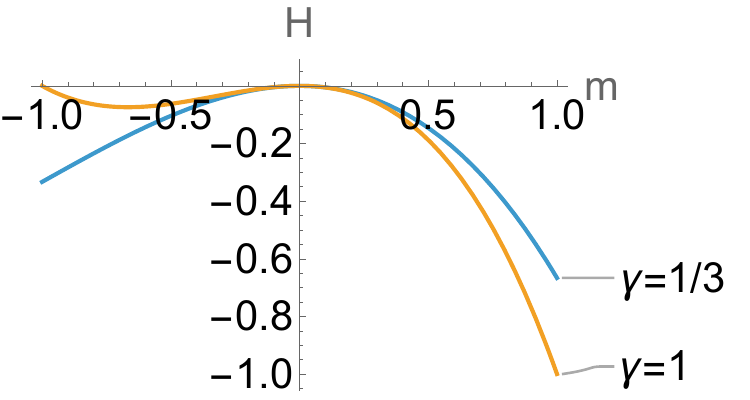}
\caption{The energy per spin, $\mathcal{H}=-\frac{1}{2}(1+\gamma m)m^2$, versus magnetization for the FIM at zero field, is plotted for two values of the feedback parameter $\gamma$. The local minimum exists only for larger value, $\gamma=1$; generally, it exists when $\gamma>\frac{2}{3}$.}
\label{Fig:energy}
\end{figure}

Here, we consider the FIM at  $h=0$; the $h\neq0$ case is similar. A super-stable MP, denoted by $M_1$, is a non-trivial local minimum of the Hamiltonian. This local minimum exists when the feedback parameter is sufficiently large, $\gamma>\frac{2}{3}$, and it reads $M_1=\frac{N}{2}(1+m_-)$ with $m_-=-\frac{2}{3\gamma}$. These assertions follow from the formula $\mathcal{H}=-\frac{1}{2}(1+\gamma m)m^2$ for the energy per spin in the FIM at zero field, cf. Eq.~\eqref{energy:FIM}; see Fig.~\ref{Fig:energy} for illustration.

At $T=0$, the flipping rates depend only on $M$ and whether it is larger or smaller than $M_1$. Suppose the initial number $M_0$ of plus spins exceeds $M_1$. More precisely, let us assume that the initial magnetization lies in the range $(m_-,0)$, equivalently   $M_1<M_0<\frac{N}{2}$.  In this case, the only possible moves are $M\to M-1$. The move $M\to M-1$ occurs at rate $M$ since there are $M$ plus spins and only such spins are flippable. The flip of a minus spin would increase energy, so it is impossible at zero temperature. Figure \ref{Fig:energy} suggests to think about $m$ as the coordinate of an analog-particle in potential $\mathcal{H}(m)=-\frac{1}{2}(1+\gamma m)m^2$: The analog-particle starting between the minimum and the maximum of the potential, $m_-<m_0<0$, runs down towards the minimum. (In contrast to a Newtonian particle that would be oscillating around the minimum, our analog-particle comes to rest upon reaching $M_1$. Furthermore, the driving `force' is not the derivative of the potential but just $M$.) 

Thus, the distribution $P_M(t)$ is non-trivial in the interval $M_1\leq M\leq M_0$. The probability to keep the maximal magnetization decays according to 
\begin{subequations}
\begin{equation}
\label{M0-eq}
\frac{dP_{M_0}}{dt} = - M_0 P_{M_0}
\end{equation}
In the bulk ($M_1<M<M_0$)
\begin{equation}
\label{M-eq}
\frac{dP_{M}}{dt}  = (M+1)P_{M+1}- M P_{M}
\end{equation}
The MP is an absorbing boundary 
\begin{equation}
\label{M1-eq}
\frac{dP_{M_1}}{dt} = (M_1+1)P_{M_1+1}
\end{equation}
\end{subequations}
These recurrent equations are solvable. The solution subject to the initial condition $P_M(0)=\delta_{M,M_0}$ reads 
\begin{equation}
\label{PM-sol}
P_M(t)=\binom{M_0}{M}\,e^{-tM}\left(1-e^{-t}\right)^{M_0-M}
\end{equation}
for $M_1<M\leq M_0$. We guessed \eqref{PM-sol} by solving \eqref{M0-eq} and then solving a few following equations \eqref{M-eq}. One then proves the validity of the solution  \eqref{PM-sol} by induction. 

The probability to be at $M=M_1$ is found by integrating \eqref{M1-eq} subject to $P_{M_1}(0)=0$, with $P_{M_1+1}$ extracted from \eqref{PM-sol}. One finds
\begin{equation}
\label{PM1-sol}
P_{M_1}(t) =\frac{B(1-e^{-t}; M_0-M_1, M_1+1)}{B(M_0-M_1, M_1+1)}
\end{equation}
where 
\begin{equation}
\label{incomplete:beta}
B(x; a, b)=\int_0^x dy\,y^{a-1} (1-y)^{b-1}
\end{equation}
is the incomplete beta function and $B(a,b)\equiv B(1; a, b)$ is the beta function. 

\begin{figure}
 \centering
    \includegraphics[width=0.46\textwidth]{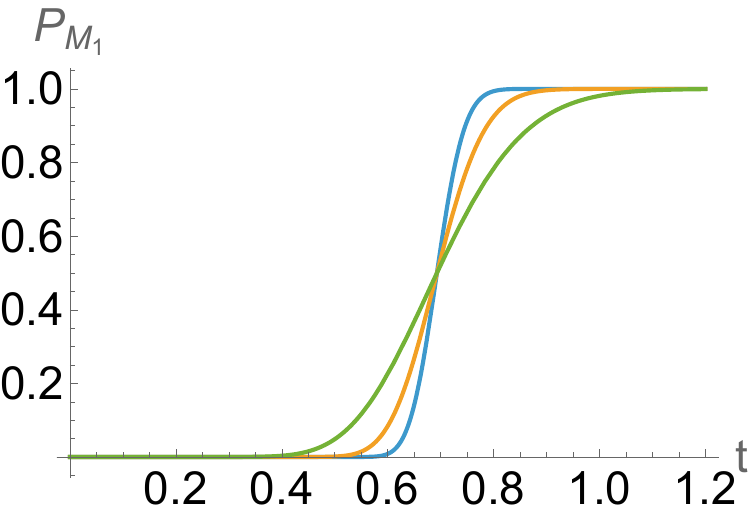}
\caption{The evolution of the probability $P_{M_1}(t)$ of being in the absorbing state. We chose $M_0=2M_1+1$. In this case, the adsorption probability is \eqref{PM1-sym}. For the three shown curves, $M_1+1=30, 100, 300$. All curves intersect at $t=\log 2$ where $P_{M_1} =\frac{1}{2}$. The front around $t=\log 2$ becomes sharper as $M_1$ increases.}
\label{Fig:PM1}
\end{figure}

In Fig.~\ref{Fig:PM1}, we plot the adsorption probability in the particular case of $M_0=2M_1+1$ when \eqref{PM1-sol} becomes
\begin{equation}
\label{PM1-sym}
P_{M_1}(t) =\frac{B(1-e^{-t}; M_1+1, M_1+1)}{B(M_1+1, M_1+1)}
\end{equation}
In this symmetric case, $P_{M_1}(t) =\frac{1}{2}$ at $t=\log 2$ independently of $M_1$. Indeed, the integral representation \eqref{incomplete:beta} gives $B(\frac{1}{2}; a, a)=\frac{1}{2}B(1; a, a)=\frac{1}{2}B(a, a)$. The width of the front around $t=\log 2$ decays as $M_1^{-1/2}$. This estimate also immediately follows from  the integral representation \eqref{incomplete:beta} of the incomplete beta function. This quantifies the notion of super-stability \cite{ma2025mixed} in the symmetric case: When $M_0=2M_1+1$ and $M_1\to \infty$, the MP state $M_1$ is reached in a time $t=\log 2$. 

We chose the symmetric case as a concrete illustration. Generally $M_0=\frac{N}{2}(1+m_0)$ and $M_1=\frac{N}{2}(1+m_-)$ with $m_-=-\frac{2}{3\gamma}$ when $\gamma>\frac{2}{3}$, and the initial magnetization obeying $m_- < m_0<0$. The integrand in the incomplete beta function  in \eqref{PM1-sol} is $y^{M_0-M_1-1}(1-y)^{M_1}$. The integrand reaches a very sharp (recall that $N\gg 1$) maximum at $y_*=\frac{m_0-m_-}{1+m_0}$, from which we conclude that the MP state $M_1$ is reached in a time
\begin{equation}
t = \log\frac{1+m_0}{1+m_-}
\end{equation}

The complementary case when $0<M_0<M_1$ can be treated similarly. The mechanical analogy suggests that only  $M\to M+1$ moves are possible as long as $M<M_1$, and $M_1$ is again an adsorbing state.  The rate of the $M\to M+1$ move is $N-M$, the number of minus spins; plus spins cannot flip. The distribution $P_M(t)$ is non-trivial in the interval $M_0\leq M\leq M_1$. The evolution equations for the probabilities $P_M(t)$ are linear and recurrent, and they admit a solution resembling \eqref{PM-sol}--\eqref{PM1-sol}.

\section{Fokker-Planck equation}
\label{LTR}

In this section, we first reduce the discrete master equation of a general mean-field IM to the continuous FP equation. Then, we use the FP equation to show that the families of equilibrium distributions near critical points are non-Gaussian, and we use such distributions to calculate transition rates between two stable equilibria. Our main goal is to reveal how the salient features on the phase diagram found in Sec.~\ref{FIM1} including temperature-induced bistability and Maxwell temperature affect equilibrium distributions and transition rates in the FIM.

We use the continuous variable $x=(m+1)/2=M/N$, $x \in [0,1]$ and rewrite the flipping rates (\ref{eq:wpmu}) as 
\begin{subequations}
\begin{align}
w_{-}(x) &= \epsilon \text{Prob} \{ x \to x-\epsilon \}= \frac{x}{2}\, [1 - f(x)]\\
w_{+}(x) &= \epsilon \text{Prob} \{ x \to x+\epsilon \}= \frac{1- x}{2}\,[1 +  f(x)]
\end{align}
\end{subequations}
where $\epsilon=N^{-1}$ and $f(x)$ denotes the tanh term. For instance, $f(x)=\tanh(\beta(h+2x-1))$ for the Curie-Weiss model, which we also call the classical IM, and $f(x)=\tanh(\beta(h+2x-1+\frac{3}{2}\gamma (2x-1)^2))$ for the FIM.

\subsection{Fokker-Planck equation and approximation error}
\label{sec:FPE-appr}

The master equation (\ref{MasterEq}) can be  rewritten as
\begin{align}
\begin{split}\label{master}
\epsilon\frac{\partial \rho(x,t)}{\partial t}=&w_{-} (x+\epsilon) \rho(x+\epsilon, t) - w_{-}(x)\rho(x, t) \\
+& w_{+} (x-\epsilon) \rho(x-\epsilon, t) - w_{+}(x)\rho(x, t)
\end{split}
\end{align}
where $\rho(x, t)=P(M, t)$. Expanding in powers of $\epsilon$ we obtain the FP equation
\begin{equation}
\partial_t \rho = \partial_x [a(x) \rho] +  \epsilon \partial_x^2 [b(x) \rho]
\end{equation}
where 
\begin{equation}
a(x)= w_{-}(x) - w_{+}(x), \quad  b(x)=\frac{w_{-}(x) + w_{+}(x)}{2}
\end{equation}
Hereinafter, we use the variable $m\equiv 2x-1$, and shortly write $\rho$ instead of $\rho(m,t)$. The FP equation in this case reads 
\begin{equation}
\label{FP}
\partial_t \rho = \partial_m [a(m) \rho] +  \epsilon \partial_m^2 [b(m) \rho]
\end{equation}
where %{\bf YPM: I have removed the denominators 2 and 4 in the below expressions of $a(m)$ and $b(m)$ since $\partial/\partial x=2\partial/\partial m$ and $\partial^2/\partial x^2=4\partial^2/\partial m^2$.}
\begin{align}
\label{ab:def}
\begin{split}
&a(m)=m-f,  \quad b(m) =1 -fm > 0,
\end{split}
\end{align}
and $f$ again denotes the tanh term but as a function of $m$, e.g., $f=\tanh[\beta (h+m+\tfrac{3}{2}\gamma m^2)]$ for the FIM.

If we drop the diffusion term, the FP equation reduces to a first-order hyperbolic partial differential equation (PDE). Such PDEs are tractable \cite{PDE-intro,Bressan} by the method of characteristics, which in the present case leads to the dynamical system
\begin{equation}
\label{dynsys}
\frac{dm}{dt}= -a(m). 
\end{equation}
For the classical IM and the FIM, this dynamical system reduces to the ODE (\ref{eq:dmdt-general}) for magnetization, and the set of equilibria is given by $a(m)=0$. 

Generally, the FP equation \eqref{FP} is a challenging PDE, albeit the equilibrium density $\rho_\text{eq}(m)$ can be readily found. Indeed, $\rho_\text{eq}(m)$ satisfies the following ODE:
\begin{equation}
\label{FP-inf}
 a(m) \rho_\text{eq}(m) +  \epsilon\, \frac{d}{dm}\,[b(m) \rho_\text{eq}(m)]=0
\end{equation}
from which
\begin{equation}
\label{FPeq}
-\epsilon \log[b(m)\rho_\text{eq}(m)] = \int_0^m \frac{a(u)}{b(u)}du + \text{const}.
\end{equation}

The FP equation as a continuous approximation to a discrete birth-death process like the master equation holds only on a codimension-1 subset of the parameter space \cite{doering2005extinction}. Here, we define the approximation error as the difference between the equilibrium distribution predicted by the master equation in Eq.~(\ref{eq:MEeq}) and that predicted by the FP equation in Eq.~(\ref{FPeq}). If we denote $r\equiv\frac{w_-(u)}{w_+(u)}$, then the integrand in Eq.~(\ref{eq:MEeq}) is $\frac{1}{2}\log(r)$, while that in Eq.~(\ref{FPeq}) is $\frac{r-1}{r+1}$. For $r>0$, these two functions intersect only at $r=1$, and their difference is
\begin{equation}
\label{eq:app-err}
    \frac{\log(r)}{2} - \frac{r - 1}{r + 1} = \frac{(r - 1)^3}{24} - \frac{(r - 1)^4}{16}+\ldots
\end{equation}
The condition $|r-1|\ll1$ implies that the integrand in Eq.~(\ref{FPeq}) must be small for the FP equation to hold. Thus, we define the continuum limit as $|a(m)|\ll1$ implying proximity to an equilibrium $m^{(0)}$ and $b(m)\to1-m^2=O(1)$ implying distance from the two aligned states $m=\pm1$. Let us write the former as $|\mu|\ll1$ where $\mu\equiv m-m^{(0)}$. Then, $a(m)=O(\mu^p)$ where the exponent $p$ can be 1, 2, or 3 depending on the nature of $m^{(0)}$ as detailed in Section \ref{sec:non-gau}. Since the interval of integration has length $O(\mu)$, the integral in Eq.~(\ref{FPeq}) is $O(\mu^{p+1})$, and the approximation error for this integral is $O(\mu^{3p+1})$ using Eq.~(\ref{eq:app-err}).

Since $b(m)=O(1)$, we can simplify Eq.~(\ref{FPeq}) to
\begin{equation}\label{eq:FP-sim}
    -\epsilon\log\rho_\text{eq}(m)\to U(m)=\int^m\frac{a(u)}{b(u)}du.
\end{equation}
Thus, $U(m)$ is effectively the free energy of the FP equation that controls the equilibrium distribution $\rho_\text{eq}(m)$.

\subsection{Equilibrium distributions near critical points}
\label{sec:non-gau}

Now we use the FP equation to predict equilibrium distributions $\rho_\text{eq}(m)$ near an equilibrium $m^{(0)}$ for the classical IM and the FIM. For an equilibrium $m^{(0)}$ at a generic choice of $T$ and $h$, we expand to leading order:
\begin{equation}
    a(m)\to\sigma\mu\quad\text{where}\quad\sigma=a'(m^{(0)}).
\end{equation}
Moreover, $b(m)\to b(m^{(0)})$, so Eq.~(\ref{eq:FP-sim}) implies that the equilibrium distribution near $m^{(0)}$ is Gaussian:
\begin{equation}
    -\epsilon\log\rho_\text{eq}(m)\to U(m)=\frac{\sigma}{2b(m^{(0)})}\mu^2.
\end{equation}

The FP equation becomes more interesting when two or more equilibria are close. On the $(T,h)$-plane, this occurs at a critical point, i.e., a bifurcation point, hereafter denoted by $(T,h)=(T^{(0)},h^{(0)})$. As shown in Section \ref{sec:CW}, the equilibrium distribution is non-Gaussian at the critical point. Now we use the FP equation to extend such distributions to a neighborhood of this critical point. Here, we have $b(m,T,h)\to b(m^{(0)},T^{(0)},h^{(0)})\equiv b^{(0)}$, while the expansion of $a(m)$ relates to normal forms of steady state bifurcations \cite{crawford1991introduction}.

\begin{figure}
    \centering
    \subfigure[]{\includegraphics[width=0.2\textwidth]{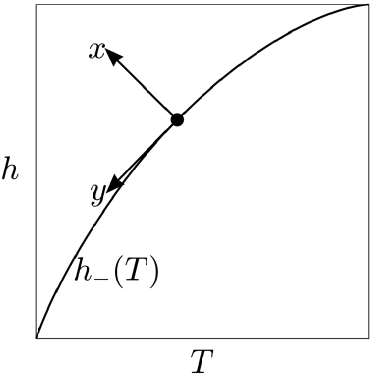}\label{fig:FIM_xy_g0m_kp}}
    \subfigure[]{\includegraphics[width=0.2\textwidth]{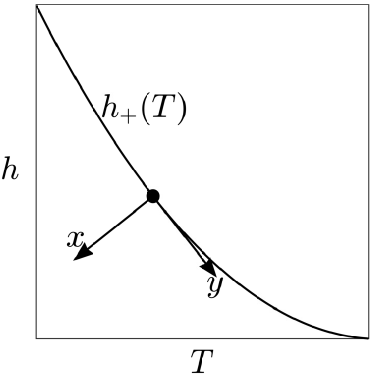}\label{fig:FIM_xy_g1p_km}}
    \subfigure[]{\includegraphics[width=0.2\textwidth]{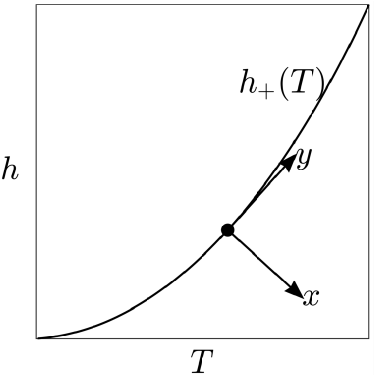}\label{fig:FIM_xy_g1p_kp}}
    \caption{Possible orientations of local coordinates based at a generic critical point $(T^{(0)},h^{(0)})$. Comparison with Fig.~\ref{fig:FIM_Th} reveals that orientations (a) \& (b) exist in both the classical IM and the FIM, while orientation (c) exists only in the FIM.}
    \label{fig:FIM_xy_crit}
\end{figure}

First, we consider a generic critical point $(T^{(0)},h^{(0)})$ where $0<T^{(0)}<T_c$ and $h^{(0)}=h_s(T^{(0)})$, $s=+$ or $-$. These critical points exist on the critical curve, which is a codimension-1 subset of the $(T,h)$ half-plane. We introduce local coordinates $(x,y)$ based at $(T^{(0)},h^{(0)})$ such that the $x$-direction is along the normal pointing into the bistable region and the $y$-direction is along the tangent with slope $k\equiv h_s'(T^{(0)})$. As shown in Fig.~\ref{fig:FIM_xy_crit}, explicitly these local coordinates are
\begin{align}
\begin{split}\label{eq:xy-crit}
x &= k(T - T^{(0)}) - (h - h^{(0)}),\\
y &= (T - T^{(0)}) + k(h - h^{(0)}).
\end{split}
\end{align}
In these local coordinates, the normal form is
\begin{equation}\label{eq:a-codim-1}
    a(m,x,y)\to\alpha x+\beta_1y^2+\beta_2y\hatmz+\sigma\hatmz^2,
\end{equation}
where  $\hatmz=m-m^{(0)}$ and 
\begin{equation*}
    \alpha=\frac{\partial a}{\partial x},\quad\beta_1=\frac{1}{2}\frac{\partial^2a}{\partial y^2},\quad
    \beta_2=\frac{\partial^2a}{\partial m\partial y},\quad\sigma=\frac{1}{2}\frac{\partial^2a}{\partial m^2},
\end{equation*}
with all derivatives evaluated at $(m,x,y)=(m^{(0)},0,0)$. If $x=O(\mu^2)$ and $y=O(\mu)$, then all terms are of the same order and $a(m,x,y)=O(\mu^2)$. Using Eq.~(\ref{eq:FP-sim}), we obtain the generic 2-parameter family of non-Gaussian distributions near the critical point:
\begin{align}
\begin{split}\label{eq:rho-codim-1}
    &-\epsilon\log\rho_\text{eq}(m) \to U(m,x,y)\\
    &=\frac{1}{b^{(0)}}\biggl((\alpha x+\beta_1y^2)\hatmz+\frac{\beta_2y}{2}\hatmz^2+\frac{\sigma}{3}\hatmz^3\biggr).
\end{split}
\end{align}

For a transverse perturbation to a critical point, we have $x=O(\mu^2)$ and $y=O(\mu^2)$. Then, considering only the leading terms, we obtain that Eq.~(\ref{eq:a-codim-1}) becomes
\begin{equation}
    a(m,x)\to\alpha x+\sigma\hatmz^2.
\end{equation}
In this case, Eq.~(\ref{dynsys}) agrees with the normal form of fold bifurcation, and we obtain the generic 1-parameter family of non-Gaussian distributions near the critical point:
\begin{equation}\label{eq:rho-fold}
    -\epsilon\log\rho_\text{eq}(m)\to U(m,x)=
    \frac{1}{b^{(0)}}\left(\alpha x\hatmz+\frac{\sigma}{3}\hatmz^3\right).
\end{equation}
The $x$-direction can only be in the second or third quadrant in the classical IM, but it can be in the second, third, or fourth quadrant in the FIM; compare Fig.~\ref{fig:FIM_Th} \& Fig.~\ref{fig:FIM_xy_crit}. Indeed, the $x$-direction being in the first or fourth quadrant is equivalent to temperature-induced bistability, which is a unique feature of the FIM.

For a tangent perturbation to a critical point, we have $x=0$ and $y=O(\mu)$. 
Keeping only the contributions of the leading order, we obtain that
 Eq.~(\ref{eq:a-codim-1}) becomes
\begin{equation}
    a(m,y)\to\beta_1y^2+\beta_2y\hatmz+\sigma\hatmz^2.
\end{equation}
If the bistable region is locally non-convex, i.e., both positive and negative $y$-directions pointing into the bistable region, then the bifurcation diagram of equilibria given by $a(m,y)=0$ consists of two intersecting lines exchanging stability. In this case, Eq.~(\ref{dynsys}) is equivalent to the normal form of transcritical bifurcation, and we obtain a non-generic 1-parameter family of non-Gaussian distributions near a critical point:
\begin{align}\label{eq:rho-trans}
\begin{split}
    -\epsilon\log\rho_\text{eq}(m) &\to U(m,y)\\
    &=\frac{1}{b^{(0)}}\left(\beta_1y^2\hatmz+\frac{\beta_2y}{2}\hatmz^2+\frac{\sigma}{3}\hatmz^3\right).
\end{split}
\end{align}
Although this family is non-generic, it is naturally realized in the FIM near the critical point $(T^{(0)},h^{(0)})=(1,0)$ with $y=T-1$; see Fig.~\ref{fig:FIM_Th}. Remarkably, Fig.~\ref{fig:FIM_Th} shows that local non-convexity holds everywhere except at the cusp in both the classical IM and the FIM. 

Assuming local non-convexity, we see that the 2-parameter expansion in Eq.~(\ref{eq:a-codim-1}) and the associated distributions in Eq.~(\ref{eq:rho-codim-1}) combine the transverse and tangent perturbations and describe universal unfolding of transcritical bifurcation. Even if the $(x,y)$-coordinates are rotated, this universal unfolding still applies to the sequence of bifurcation diagrams on the $(y,m)$-plane as $x$ varies. This universal unfolding is naturally realized in the FIM near the critical point $(T^{(0)},h^{(0)})=(1,0)$ with $x=-h$ and $y=T-1$ as discussed in Section \ref{sec:FIM-non}.

\begin{figure}
    \centering
    \subfigure[]{\includegraphics[width=0.2\textwidth]{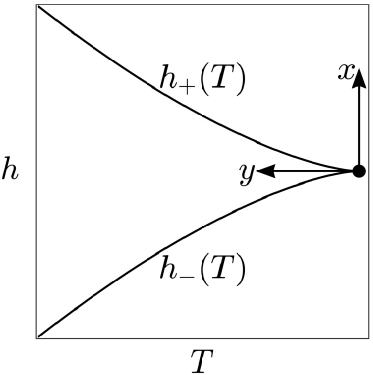}\label{fig:FIM_xy_g0_cusp}}
    \subfigure[]{\includegraphics[width=0.2\textwidth]{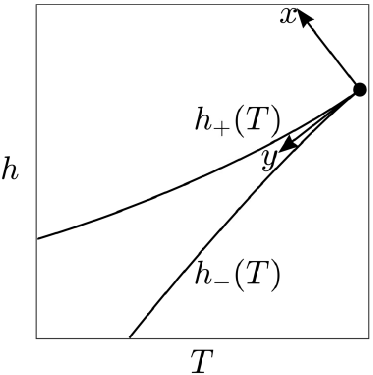}\label{fig:FIM_xy_g1_cusp}}
    \caption{Possible orientations of local coordinates based at the cusp $(T_c,h_c)$. Comparison with Fig.~\ref{fig:FIM_Th} reveals that orientation (a) applies to the classical IM, while orientation (b) applies to the FIM.}
    \label{fig:FIM_xy_cusp}
\end{figure}

Next, we consider the codimension-2 critical point, i.e., the cusp $(T^{(0)},h^{(0)})=(T_c,h_c)$. Here, since the equilibrium was denoted by $m^{(0)}=m_c$, we rename $b^{(0)}\equiv b_c$ for consistency. We introduce local coordinates $(x,y)$ based at $(T_c,h_c)$ such that the $y$-direction is along the center of the wedge, i.e., the Maxwell curve $h_*(T)$, and the $x$-direction is perpendicular to the $y$-direction; see Fig.~\ref{fig:FIM_xy_cusp}. For the FIM, since the cusp satisfies $T'(m_c)=h'(m_c)=0$ with $T(m)$ and $h(m)$ given by Eq.~\eqref{eq:Tm_hm}, the local slope of the Maxwell curve is
\begin{equation}
    k\equiv h_*'(T_c)=\frac{h''(m_c)}{T''(m_c)},
\end{equation}
and the local coordinates are
\begin{align}
\begin{split}\label{eq:xy-cusp}
x &= -k(T - T^{(0)}) + (h - h^{(0)}),\\
y &= -(T - T^{(0)}) - k(h - h^{(0)}).
\end{split}
\end{align}
In these local coordinates, the normal form is
\begin{equation}\label{eq:a-codim-2}
    a(m,x,y)\to\alpha_1x+\alpha_2y\hatmc+\sigma\hatmc^3,
\end{equation}
where $\hatmc=m-m_c$ and 
\begin{equation*}
    \alpha_1=\frac{\partial a}{\partial x},\quad\alpha_2=\frac{\partial^2a}{\partial m\partial y},\quad
    \sigma=\frac{1}{6}\frac{\partial^3a}{\partial m^3},
\end{equation*}
with all derivatives evaluated at $(m,x,y)=(m_c,0,0)$. If $x=O(\mu^3)$ and $y=O(\mu^2)$, then all terms are on the same order and $a(m,x,y)=O(\mu^3)$ overall. Using Eq.~(\ref{eq:FP-sim}), we obtain the generic 2-parameter family of non-Gaussian distributions near the cusp:
\begin{align}
\begin{split}\label{eq:rho-codim-2}
    &-\epsilon\log\rho_\text{eq}(m)\to U(m,x,y)\\
    &=\frac{1}{b_c}\left(\alpha_1x\hatmc+\frac{\alpha_2y}{2}\hatmc^2+\frac{\sigma}{4}\hatmc^4\right).
\end{split}
\end{align}
This generalizes the 1-parameter family of non-Gaussian distributions (\ref{crit-window}) in the scaling window to general mean-field IMs with a nonzero external magnetic field. 

For a transverse perturbation to the cusp, $x=O(\mu^3)$ and $y=O(\mu^3)$, so the second term in Eq.~(\ref{eq:a-codim-2}) drops out, which implies no bifurcations. For a tangent perturbation to the cusp, $x=0$ and $y=O(\mu^2)$, so Eq.~(\ref{eq:a-codim-2}) becomes
\begin{equation}
    a(m,y)\to\alpha_2y\hatmc+\sigma\hatmc^3.
\end{equation}
In this case, Eq.~(\ref{dynsys}) agrees with the normal form of pitchfork bifurcation. Thus, a non-generic 1-parameter family of non-Gaussian distributions near the cusp reads
\begin{equation}\label{eq:rho-pitch}
    -\epsilon\log\rho_\text{eq}(m)\to U(m,y)=
    \frac{1}{b_c}\left(\frac{\alpha_2y}{2}\hatmc^2+\frac{\sigma}{4}\hatmc^4\right).
\end{equation}
Although this family is non-generic, it is naturally realized in the classical IM near the cusp $(T_c,h_c)=(1,0)$ with $y=1-T$; compare Fig.~\ref{fig:FIM_Th} \& Fig.~\ref{fig:FIM_xy_cusp}. In this case, Eq.~(\ref{eq:rho-pitch}) agrees with Eq.~(\ref{crit-window}) for the equilibrium distributions in the scaling window.

The 2-parameter generalized quartic potential $U(m,x,y)$ in Eq.~(\ref{eq:rho-codim-2}) is well known in Landau's theory of phase transitions, while Eq.~(\ref{dynsys}) with the 2-parameter expansion in Eq.~(\ref{eq:a-codim-2}) agrees with the normal form of cusp bifurcation \cite{guckenheimer2007cusp}. This bifurcation is based on cusp catastrophe, which has been extensively studied in catastrophe theory \cite{thom2018structural,zeeman2006classification}. As $x$ varies, the sequence of bifurcation diagrams on the $(y,m)$-plane near the cusp is topologically equivalent to the sequence for the classical IM as shown in Fig.~\ref{fig:IM_bif_T}. However, if the $(x,y)$-coordinates are rotated, then the sequence near the cusp is topologically equivalent to the sequence for the FIM with $\gamma\in(0,\gamma_*)$ as shown in Fig.~\ref{fig:FIM_bif_T_g23}. Thus, the FIM provides a minimal model of a rotated cusp. This guarantees Maxwell temperatures nearby since the Maxwell curve bifurcates from the cusp.

\subsection{Transition time near critical points}

We can use non-Gaussian distributions near critical points in Section \ref{sec:non-gau} to predict transition times between two stable equilibria. First, we consider non-Gaussian distributions near a generic critical point $(T^{(0)},h^{(0)})$ where $0<T^{(0)}<T_c$ and $h^{(0)}=h_s(T^{(0)})$, $s=+$ or $-$. The transition rate $C_{+-}$ depends on the transition path $m\in[m_0,m_+]$, which is short when $h^{(0)}=h_-(T^{(0)})$. Similarly, the transition rate $C_{-+}$ depends on the transition path $m\in[m_-,m_0]$, which is short when $h^{(0)}=h_+(T^{(0)})$. In either case, a non-Gaussian distribution applies to a tubular neighborhood of the critical curve and covers the transition path from the less probable state to the more probable state, but not the reverse.

The 1-parameter family of non-Gaussian distributions in Eq.~(\ref{eq:rho-fold}) describes the fold bifurcation. In this case, a stable equilibrium and a saddle coexist when $\alpha \sigma x < 0$. This condition must be equivalent to $x>0$ since the $x$-direction points into the bistable region, so we have $\alpha\sigma<0$. The transition rate $C_{+-}$ or $C_{-+}$ is given by the potential difference between these two equilibria:
\begin{equation}\label{eq:C-fold}
    \mu = \pm \sqrt{-\frac{\alpha x}{\sigma}}\quad\Rightarrow\quad C = \frac{4}{3b^{(0)}} \cdot \frac{(-\alpha\sigma x)^{3/2}}{\sigma^2}.
\end{equation}
Thus, the transition occurs only for $x>0$, and the transition rate scales as $C\to x^\frac{3}{2}$ with exponent $\frac{3}{2}$.

The 1-parameter family of non-Gaussian distributions in Eq.~(\ref{eq:rho-trans}) describes transcritical bifurcation when local non-convexity $\Delta_0\equiv\beta_2^2 - 4 \beta_1 \sigma > 0$ holds. In this case, a stable equilibrium and a saddle coexist when $y\neq0$, and the transition rate $C_{+-}$ or $C_{-+}$ is given by the potential difference between these two equilibria:
\begin{equation}\label{eq:C-trans}
    \mu = \frac{-\beta_2 \pm \sqrt{ \Delta_0 }}{2\sigma} \cdot y
    \quad\Rightarrow\quad C = \frac{1}{6b^{(0)}} \cdot \frac{\Delta_0^{3/2} |y|^3}{ \sigma^2 }.
\end{equation}
Thus, the transition occurs for both signs of $y$, and the transition rate scales as $C\to|y|^3$ with exponent 3.

Now we consider the 2-parameter family of non-Gaussian distributions in Eq.~(\ref{eq:rho-codim-1}). If the discriminant $\Delta(x,y)\equiv-4 \alpha \sigma x + \Delta_0 y^2 > 0$, then a stable equilibrium and a saddle coexist:
\begin{equation}
    \mu = \frac{-\beta_2 y \pm \sqrt{\Delta(x,y)}}{2 \sigma}.
\end{equation}
The critical curve is $\Delta(x,y)=0$, i.e., $x=\Delta_0y^2/(4\alpha\sigma)$, with bistability for $x>0$. If local non-convexity $\Delta_0 > 0$ holds, then the coexistence condition is $|y|>2\sqrt{\alpha\sigma x/\Delta_0}$ when $x<0$ and $y\in\mathbb{R}$ when $x>0$. Thus, the bifurcation diagram on the $(y,m)$-plane exhibits universal unfolding of transcritical bifurcation as $x$ varies. However, if local convexity $\Delta_0 < 0$ holds, then the coexistence condition is nil when $x<0$ and $|y|<2\sqrt{\alpha\sigma x/\Delta_0}$ when $x>0$. Thus, the bifurcation diagram on the $(y,m)$-plane exhibits creation of an isola, i.e., a loop with two folds, as $x$ increases. There are no isolas in the classical IM or the FIM since local non-convexity persists as shown in Fig.~\ref{fig:FIM_Th}. The transition rate $C_{+-}$ or $C_{-+}$ is again given by the potential difference between the two equilibria:
\begin{equation}
    C = \frac{[\Delta(x,y)]^{3/2}}{6 b^{(0)} \sigma^2}.
\end{equation}
This agrees with Eq.~(\ref{eq:C-fold}) when $y=0$ and Eq.~(\ref{eq:C-trans}) when $x=0$, so the discriminant $\Delta(x,y)$ measures the effective distance between the point $(x,y)$ and the critical curve.

Next, we consider non-Gaussian distributions near the cusp $(T_c,h_c)$. The transition paths for the two rates $C_{+-}$ and $C_{-+}$ can be combined into $m\in[m_-,m_+]$. Thus, a non-Gaussian distribution applies to the wedge $h_-(T)<h<h_+(T)$ with $T<T_c$ and covers the transition path in either direction. Such transitions were originally studied using the classical theory of WKB approach and called Kramers transitions. The general theory of large deviations, which works for general gradient-like dynamical systems with small random perturbations, was later developed as Freidlin-Wentzell theory \cite{Freid}.  

The 1-parameter family of non-Gaussian distributions in Eq.~(\ref{eq:rho-pitch}) describes a pitchfork bifurcation. In this case, two stable equilibria $m_\pm$ and a saddle $m_0$ coexist when $\alpha_2 y < 0$ since $\sigma>0$. This must be equivalent to $y>0$ since the $y$-direction is along the center of the wedge, so we have $\alpha_2<0$ by definition. The transition rates $C_{+-}$ and $C_{-+}$ are given by the potential barriers between $m_\pm$, i.e., the potential differences between $m_\pm$ and $m_0$:
\begin{equation}
\begin{split}
    &m_\pm = m_c \pm \sqrt{ -\frac{\alpha_2 y}{\sigma} },\quad m_0 = m_c \quad\Rightarrow \\
    &C_{+-}=C_{-+}=U(m_0)-U(m_\pm)=\frac{(\alpha_2 y)^2}{4b_c\sigma}.
\end{split}
\end{equation}
Thus, the transitions occur only for $y>0$. The two rates are equal due to up-down symmetry of pitchfork bifurcation and scale as $C\to y^2$ with exponent 2.

\begin{figure}
    \centering
    \includegraphics[width=0.4\textwidth]{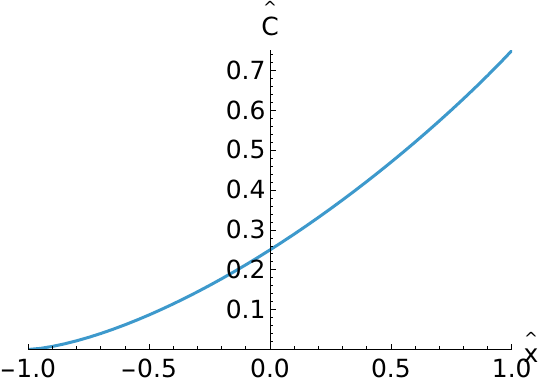}
    \caption{Plot of the  function $\hat{C}(\hat{x})$ defined by Eq.~(\ref{eq:C-hat}), which describes potential differences near a cusp.}
    \label{fig:C-hat-x-hat}
\end{figure}

Now we consider the 2-parameter family of non-Gaussian distributions in Eq.~(\ref{eq:rho-codim-2}). If the discriminant $\Delta(x,y) \equiv -4 (\alpha_2 y)^3 - 27 \alpha_1^2 \sigma x^2 > 0$, then two stable equilibria $m_\pm$ and a saddle $m_0$ coexist. The critical curve is $\Delta(x,y)=0$, i.e., $x=\left[2/\left(3\sqrt{3}\right)\right]\cdot\left[(-\alpha_2 y)^\frac{3}{2}/\left(\alpha_1\sqrt{\sigma}\right)\right]$. Thus, we introduce the rescaled variables
\begin{equation}\label{eq:cusp-hat}
    \hat{m}\equiv\frac{m-m_c}{\sqrt{-\alpha_2y/\sigma}},\qquad\hat{x}=\frac{3\sqrt{3}}{2}\cdot\frac{\alpha_1\sqrt{\sigma}x}{(-\alpha_2 y)^\frac{3}{2}}
 \end{equation}
and also rescale the potential 
\begin{equation}
\hat{U}=\frac{b_c\sigma U}{(\alpha_2y)^2}\,.
\end{equation}
In terms of the rescaled variables, the rescaled potential reads
\begin{equation}
    \hat{U}(\hat{m},\hat{x})=\frac{1}{4}\hat{m}^4-\frac{1}{2}\hat{m}^2+\frac{2}{3\sqrt{3}}\hat{x}\hat{m}.
\end{equation}
The equilibria $\hat{m}_-<\hat{m}_0<\hat{m}_+$ satisfy $\partial\hat{U}/\partial\hat{m}=0$, which is a depressed cubic equation:
\begin{equation}
    \hat{m}^3-\hat{m}+\frac{2}{3\sqrt{3}}\hat{x}=0.
\end{equation}
The bistable region is $\hat{x}\in(-1,1)$. Let us define the  function with dimensionless variables:
\begin{equation}\label{eq:C-hat}
    \hat{C}(\hat{x})\equiv\hat{U}(\hat{m}_0,\hat{x})-\hat{U}(\hat{m}_-,\hat{x}),
\end{equation}
which is  the transition rate $C_{-+}$ rescaled. The analytical expression of $\hat{C}(\hat{x})$ is complicated, but one can check that $\hat{C}(\hat{x})$ increases monotonically for $\hat{x}\in[-1,1]$; see Fig.~\ref{fig:C-hat-x-hat}. One can also show that $\hat{C}(-1)=0$, $\hat{C}(0)=\frac{1}{4}$, and $\hat{C}(1)=\frac{3}{4}$, so the transition rate triples between the Maxwell curve and the critical curve. The transition rates in dimensional variables are scaling functions
\begin{equation}
C_{+-}(x,y)=C_{-+}(-x,y)=\frac{(\alpha_2y)^2}{b_c\sigma}\,\hat{C}(\hat{x})
\end{equation}
effectively depending only on $\hat{x}$ defined in Eq.~(\ref{eq:cusp-hat}). 

\begin{table*}
    \centering
    \begin{tabular}{|c|c|c|c|c|c|}
    \hline
    Base point & $x$ & $y$ & $C(x)$ & $C(y)$ & $C(x,y)$ \\
    \hline
    Critical point & $O(\mu^2)$ & $O(\mu)$ & $O(x^\frac{3}{2})$ & $O(|y|^3)$ & $O\left((y^2+kx)^\frac{3}{2}\right)$ \\
    \hline
    Cusp & $O(\mu^3)$ & $O(\mu^2)$ & N/A & $O(y^2)$ & $O\left(y^2\hat{C}\left(kx/y^\frac{3}{2}\right)\right)$ \\
    \hline
    \end{tabular}
    \caption{Summary of scalings of local coordinates and transition rates. The second and third rows apply respectively to Fig.~\ref{fig:FIM_xy_crit} and Fig.~\ref{fig:FIM_xy_cusp}. The transition rate $C(x)$ describes a transverse perturbation and depends on $x$ only. The transition rate $C(y)$ describes a tangent perturbation and depends on $y$ only. The transition rate $C(x,y)$ describes a general perturbation depending on both $x$ and $y$, where $k$ is a constant, and $\hat{C}$ is defined in Eq.~(\ref{eq:C-hat}) and plotted in Fig.~\ref{fig:C-hat-x-hat}.}
    \label{tab:sum-sca}
\end{table*}

Overall, the FP equation works near the two critical curves, although the scalings change when these two curves meet at the cusp; see Table \ref{tab:sum-sca} for a summary of scalings of local coordinates and transition rates. The classical IM and the FIM obey the same scaling laws, since both belong to the mean-field universality class, but the FIM provides more freedom to control the directions of steady-state bifurcations.

\section{Discussion}
\label{sec:disc}

Taken together, our results show that making the pairwise interaction of the IM depend on the system's own magnetization via a linear feedback function fundamentally reorders the mean-field phase diagram and the statistics of fluctuations.

In contrast with the Curie-Weiss model, where increasing the temperature always erodes bistability, increasing the temperature can favor bistability in the FIM. This temperature-induced bistability endows the FIM with a unit-temperature transcritical bifurcation point at zero magnetic field. Moreover, at small positive magnetic fields, the phase diagram of the FIM exhibits a re-entrant bistable window with a cusp. Therefore, fixed-field slices can display one, two, or even three critical temperatures. Within the wedge of bistability, a Maxwell curve rises to the cusp and separates the dominance regions of the two phases. This yields a unique Maxwell temperature for each fixed field in a finite interval, with field increases favoring the upper-magnetization phase and temperature increases favoring the lower one.

In both the Curie-Weiss model and the FIM, the FP equation describes the dynamics near an equilibrium. Near critical points, this effective description yields explicit families of non-Gaussian equilibrium distributions that relate to the normal forms of the associated steady-state bifurcations. Moreover, these distributions can predict barrier functions and transition rates for rare inter-well transitions with lifetimes exponential in system size. In these settings, the FIM offers more freedom than the Curie-Weiss model to control the directions of steady-state bifurcations due to temperature-induced bistability.

We also note that the FIM exhibits a dynamical finite-size effect at zero temperature: for sufficiently strong feedback, a super-stable MP appears and an initial probability distribution collapses to a Dirac delta function in finite time. As shown in Ref.~\cite{ma2025mixed}, super-stability is absent in the Curie-Weiss model unless the coupling is antiferromagnetic. Thus, the FIM provides a minimal model of finite-time collapse of the probability distribution in an exactly solvable mean-field spin system.

The feedback coupling in the FIM can be generalized to an arbitrary function $f(m)$. At zero temperature, one can infer a feedback function $f$ and thus construct a FIM for any bifurcation diagram representing a binary phase transformation \cite{ma2025mixed}. Thus, FIMs can predict the fate of such bifurcation diagrams at finite temperatures without additional information. A key question is how the function $f$ controls key features of the phase diagram at finite temperatures, including critical curves, cusps, and Maxwell curves. Practically, delayed feedback can also be relevant, which may yield oscillatory regimes.

A major application of stochastic bistable systems resembling FIMs is to model rare events defined as transitions between the two stable states. Here, a major goal is to recover optimal escape paths for such transitions, also known as instantons. The classical Freidlin-Wentzell theory calculates the optimal path via the Hamiltonian dynamics of an auxiliary dynamical system \cite{Freid}. A well-known application of this theory is to computing optimal paths in chemical kinetics \cite{dykman1994large}. Later, efficient algorithms, e.g., the string method, were developed and widely used to study rare events \cite{weinan2002string}. Such methods can be readily applied to calculate optimal paths in FIMs. Near critical curves, our results in Table \ref{tab:sum-sca} generalize the classical result that the transition rate exponent is $3/2$ near saddle-nodes \cite{dykman1994large}. Thus, the optimal paths near critical curves may inherit some universality from the normal forms. In the low temperature limit, the possible super-stability of the two stable states and the super-instability of the intermediate state may yield key insights into extremely rare events.

The statistical mechanics and dynamics of systems with long-range interactions have attracted much recent interest \cite{Ruffo09,defenu2023long}. In this setting, a natural way to construct finite-dimensional FIMs is to generalize the long-range Ising model (LRIM) pioneered by Dyson \cite{Dyson69}, yielding what we call feedback LRIM. Such models contain $p$-spin interactions of the form $J_p\propto D_p^{-a_p}$, $p\geq2$, where $D_p$ is a $p$-spin distance and $a_p>0$. Thus, the feedback LRIM approach interpolates between mean-field FIMs for small $a_p$'s and nearest-neighbor FIMs for large $a_p$'s. The classical LRIM contains only the 2-spin interaction $J_2$ with $D_2$ being the geodesic distance. A classical renormalization group calculation reveals the location $a_2=a_2^*$ where the critical exponents cross over from long-range to short-range \cite{sak1973recursion}. The equilibrium properties of the classical LRIM near this crossover are still actively explored \cite{angelini2014relations,Behan_2017}. In feedback LRIMs, new types of crossovers are possible depending on the network topology and also the distance function $D_p$, $p\geq3$, yielding a rich subject to explore further.

The theory of tipping points in open systems applies to climate change, among other fields of recent interest. There are three tipping mechanisms: bifurcation, noise-induced, and rate-dependent \cite{ashwin2012tipping}. At zero temperature, FIMs in a time-varying magnetic field provide minimal models for bifurcation and rate-dependent tipping \cite{ma2025mixed}. For FIMs in the bistable regime at finite temperatures, transitions between the two stable states can be identified as noise-induced tipping.

Because a macroscopic observable feedbacks onto the microscopic coupling, the same mechanism offers a compact modeling language for systems where interactions co-evolve with context. In information ecosystems, polarization and platform amplification map naturally to this framework; large-scale evidence on faster and farther diffusion of false news and on echo-chamber formation is consistent with re-entrant polarization, heavy-tailed fluctuations near critical sets, and asymmetric narrative lifetimes predicted by feedback Ising dynamics \cite{Vosou,Cinelli}. In human--AI ecosystems, mean-field mixtures of human and model agents with pair and higher-order interactions display abrupt regime changes as composition and interaction strengths vary, making the Maxwell frontier and barrier-based lifetime scalings useful for alignment/governance diagnostics and for inverse modeling that fits feedback laws to equal-likelihood frontiers and dwell times \cite{AI-Human}. In center--periphery (empires--provinces) settings \cite{Vak25,Sud25}, allegiance plays the role of the order parameter while information and resource loops supply feedback; the coexistence of folds and a cusp implies multiple cohesion thresholds, and the Maxwell curve provides a data-driven boundary between center-loyal and provincial-dissent phases---estimable from transition statistics \cite{Castellano2009_RevModPhys}.

Temperature-induced bistability in FIMs represents a counterintuitive type of noise-induced tipping. Typically, the transition frequency increases as the temperature (noise) increases until bistability is destroyed. However, temperature-induced bistability in FIMs allows the transition frequency to decrease as the temperature increases. Thus, our mean-field model exhibits high-temperature ordered phases in a restricted sense. This defies conventional wisdom. Interestingly, ordered phases persisting in arbitrarily high temperatures have been recently reported \cite{Han2025,Zohar25,Zohar26}, and named entropic order, in quantum and classical many-body systems.

\section*{Acknowledgments}
This research was made possible by a Research-in-Groups programme funded by the International Centre for Mathematical Sciences, Edinburgh, U.K. I.S. gratefully acknowledges support from the Division of Physics at the National Science Foundation (NSF) through Grant No. PHY-2102906. S.A.V. is funded by the Russian Ministry of Science and Education (project 124040800009-8).

\appendix

\section{Derivation and solution of the Fokker-Planck equation for the Curie-Weiss model at the cusp}
\label{sec:conv-eq-zero}

In Sec.~\ref{sec:FPE-appr}, we discussed when the FP equation provides an accurate description of the evolution of the magnetization in a general mean-field IM. Here we consider the simplest mean-field IM, the Curie-Weiss model. We focus on the most interesting critical dynamical behavior where the FP equation provides the asymptotically exact description if the initial magnetization is close to zero. The most natural initial condition is when it is zero 
\begin{equation}
\label{IC}
P(m,t=0)=\delta(m)
\end{equation}

First, we derive the FP equation; it can be extracted from the general results of Sec.~\ref{sec:FPE-appr}, but it is convenient to see the details for the Curie-Weiss model. First, we re-write the exact equations for $P(m)\equiv P_M$ as
\begin{eqnarray}
\label{Pmt}
4\epsilon\,\frac{d P(m)}{dt}&=& W_+(m) P(m+2\epsilon) + W_-(m) P(m-2\epsilon) \nonumber\\
&-&2 W_0(m)P(m)
\end{eqnarray}
Here $\epsilon=1/N$ as before and we suppress the time variable, so $P(m)\equiv P(m,t)$. The rates are
\begin{equation}
\label{W-rates}
\begin{split}
&W_+(m) = (1+m+2\epsilon)[1-\tanh(m+2\epsilon)]\\
&W_-(m)  = (1-m+2\epsilon)[1+\tanh(m-2\epsilon)]\\
& W_0(m) = 1-m \tanh m
\end{split}
\end{equation}
We are interested in the large $N$, equivalently small $\epsilon$ behavior. The magnetization $m$ is also small, viz., of the order of $N^{-1/4}$ in the long time regime, cf. \eqref{crit-inf}, and even smaller at the intermediate times. [Recall that the initial magnetization is zero, \eqref{IC}.] Since $\epsilon\ll 1$ and $m\ll 1$, we can expand the rates \eqref{W-rates}:
\begin{equation*}
\begin{split}
&W_+ = 1-(m+2\epsilon)^2+\frac{(m+2\epsilon)^3}{3}+\ldots\\
&W_- = 1-(m-2\epsilon)^2-\frac{(m-2\epsilon)^3}{3}+\ldots\\
&W_0 = 1 - m^2 +\ldots
\end{split}
\end{equation*}
Substituting these expansions into \eqref{Pmt}, replacing  differences by derivatives, and taking into account that $\epsilon\ll m$, we arrive at
\begin{equation}
\label{FPE}
\frac{\partial P}{\partial t}  = \frac{1}{N} \frac{\partial^2 P}{\partial m^2} + \frac{1}{3} \frac{\partial}{\partial m}(m^3 P)
\end{equation}
in the leading order. Equation \eqref{FPE} can be interpreted as a FP equation for the Brownian particle with diffusion coefficient $D=\frac{1}{N}$ in a quartic potential.

In the $t\to\infty$ limit, the probability distribution becomes stationary, so it obeys
\begin{equation}
0=\frac{1}{N} \frac{d^2 P_{\rm eq}}{d m^2} + \frac{1}{3}  \frac{d}{d m}(m^3 P_{\rm eq})
\end{equation}
whose solution is indeed given by \eqref{crit-inf}.

Before presenting the exact solution of \eqref{FPE} it we note that diffusion dominates in the earlier time regime where the solution is close to the standard Gaussian distribution
\begin{equation}
\label{Gauss}
P(m,t)\simeq \sqrt{\frac{N}{4\pi t}}\,\exp\!\left[-\frac{Nm^2}{4t}\right]
\end{equation}
The crossover time is found by comparing \eqref{crit-inf} and \eqref{Gauss} to yield $t_c\sim \sqrt{N}$. Thus \eqref{crit-inf} applies when $t\gg t_c$, while \eqref{Gauss}  holds when $t\ll t_c$.

To solve Eq.~\eqref{FPE} subject to the initial condition \eqref{IC} for all $t>0$, we perform the Laplace transform
\begin{equation}
\label{Lap:def}
\Pi(m,s) = \int_0^\infty dt\,e^{-st}P(m,t)
\end{equation}
and find
\begin{equation}
\label{Lap}
s\Pi - \delta(m) = \frac{1}{N} \frac{d^2 \Pi}{d m^2} + \frac{1}{3}  \frac{d}{d m}(m^3 \Pi)
\end{equation}
Equation \eqref{FPE} and the initial condition \eqref{IC} are symmetric under the $m\leftrightarrow -m$ reflection. Therefore, $\Pi(m,s)=\Pi(-m,s)$, and it suffices to solve \eqref{Lap} for $m>0$, i.e.,
\begin{equation}
\label{Lap+}
s\Pi = \frac{1}{N} \frac{d^2 \Pi}{d m^2} + \frac{1}{3}  \frac{d}{d m}(m^3 \Pi)
\end{equation}
The delta function in \eqref{Lap} is accounted by the boundary condition
\begin{equation}
\label{Lap:BC}
\frac{d \Pi}{d m}\Big|_{m=0}=-\frac{N}{2}
\end{equation}
Solving \eqref{Lap+}--\eqref{Lap:BC} gives
\begin{eqnarray}
\label{Lap-sol}
\Pi &=& C(s,N) \text{HeunB}\big[\tfrac{Ns}{4}, \tfrac{N}{4}, \tfrac{1}{2}, 0,  \tfrac{N}{6}; m^2\big] \nonumber \\
&-& \tfrac{1}{2} Nm \text{HeunB}\big[\tfrac{Ns}{4}, \tfrac{N}{3}, \tfrac{3}{2}, 0,  \tfrac{N}{6}; m^2\big]
\end{eqnarray}
where $\text{HeunB}[a_1,a_2,a_3,a_4,a_5; z]$ denotes the bi-confluent Heun function \cite{NIST}.

To determine the amplitude $C(s,N)$, we rely on
\begin{equation}
\label{norm}
\int_0^\infty dm\,\Pi(m,s) = \frac{1}{2s}
\end{equation}
which follows from the normalization condition
\begin{equation}
\int_0^\infty dm\,P(m,t) = \frac{1}{2}
\end{equation}
Substituting \eqref{Lap-sol} into \eqref{norm}, we fix the amplitude $C(s,N)$ in the solution \eqref{Lap-sol}
\begin{subequations}
\begin{align}
\label{amp}
&C(s,N) = \frac{s^{-1}+NH_2(s,N)}{2H_1(s,N)}\\
\label{H1}
&H_1= \int_0^\infty dm\, \text{HeunB}\big[\tfrac{Ns}{4}, \tfrac{N}{4}, \tfrac{1}{2}, 0,  \tfrac{N}{6}; m^2\big]\\
\label{H2}
&H_2= \int_0^\infty dm\, m\text{HeunB}\big[\tfrac{Ns}{4}, \tfrac{N}{3}, \tfrac{3}{2}, 0,  \tfrac{N}{6}; m^2\big]
\end{align}
\end{subequations}

Equation \eqref{amp} shows that the amplitude has a simple pole at $s=0$ with residue $\frac{1}{2 H_1(0,N)}$.
A useful identity
\begin{equation}
\label{Heun:ident}
\text{HeunB}\big[0, \tfrac{N}{4}, \tfrac{1}{2}, 0,  \tfrac{N}{6}; m^2\big] = \exp\!\left[-\frac{Nm^4}{12}\right]
\end{equation}
allows us to determine the residue:
\begin{equation}
\label{CsN}
\lim_{s\downarrow 0} s C(s,N) = CN^{1/4}
\end{equation}
with $C$ given by Eq.~\eqref{crit-C}. Alternatively, one can derive \eqref{CsN} using the definition \eqref{Lap:def} of the Laplace transform and observing that
\begin{equation}
\lim_{s\downarrow 0} s \Pi(m,s) = P_{\rm eq}(m)
\end{equation}

Thus, the Laplace transform is a combination of the bi-confluent Heun functions. It appears impossible to express the inverse Laplace transform in terms of known special functions.

\bibliography{references}

\end{document}